\documentclass[aps,prx,twocolumn,superscriptaddress,nofootinbib,floatfix]{revtex4-2}

\usepackage[T1]{fontenc}
\usepackage[utf8]{inputenc}
\usepackage{amsmath,amssymb,amsthm,mathtools,bm}
\usepackage{graphicx}
\usepackage[hidelinks]{hyperref}
\hypersetup{hypertexnames=false}
\usepackage{physics}

\newtheorem{theorem}{Theorem}
\newtheorem{proposition}{Proposition}

\newcommand{\oper}[1]{\mathbf{\mathsf{#1}}}

\begin{document}

\title{Optimal Interaction Free Localization with Multipath Interferometers}

\author{Anubhav Chaturvedi}
\affiliation{Division of Quantum Optics and Information, Institute of Theoretical Physics and Astrophysics, Faculty of Applied Physics and Mathematics, Gda\'nsk University of Technology, 80-233 Gda\'nsk, Poland}

\author{Jorge Escand\'on-Monardes}
\affiliation{International Centre for Theory of Quantum Technologies, University of Gda\'nsk, 80-309 Gda\'nsk, Poland}
\affiliation{ANID -- Millennium Science Initiative Program -- Millennium Institute for Research in Optics, Universidad de Concepci\'on, 160-C Concepci\'on, Chile}

\author{Esteban S.~G\'omez}
\affiliation{Departamento de F\'{\i}sica, Universidad de Concepci\'on, 160-C Concepci\'on, Chile}
\affiliation{ANID -- Millennium Science Initiative Program -- Millennium Institute for Research in Optics, Universidad de Concepci\'on, 160-C Concepci\'on, Chile}

\author{Gustavo Lima}
\affiliation{Departamento de F\'{\i}sica, Universidad de Concepci\'on, 160-C Concepci\'on, Chile}
\affiliation{ANID -- Millennium Science Initiative Program -- Millennium Institute for Research in Optics, Universidad de Concepci\'on, 160-C Concepci\'on, Chile}

\author{Stephen P. Walborn}
\affiliation{Departamento de F\'{\i}sica, Universidad de Concepci\'on, 160-C Concepci\'on, Chile}
\affiliation{ANID -- Millennium Science Initiative Program -- Millennium Institute for Research in Optics, Universidad de Concepci\'on, 160-C Concepci\'on, Chile}

\begin{abstract}
Interaction-free measurement (IFM) certifies the presence of an absorbing object without a photon ever being absorbed by it. When several candidate locations are available, existing protocols can also identify which one holds the absorber, but they do so by testing paths sequentially through two-path interferometers, resolving a binary presence question at each step. We propose a different approach: probing all candidate locations at once, with the photon prepared in a coherent superposition across every path before a single measurement resolves the outcome. We prove that a three-stage protocol built on a \(d\)-path interferometer attains the exact one-shot optimum for this task, and we extend it to \(k\) absorbers among \(d\) paths, where the no-absorption branch encodes the entire absorber subset coherently rather than revealing individual locations one by one. The dark port therefore ceases to be a mere witness of presence and becomes a location-resolving signal. We then move beyond single-pass strategies using the quantum-comb formalism, casting the problem as an exact optimization over all multi-pass strategies and showing that adaptive protocols surpass the one-shot ceiling. Enriching the interferometer geometry with one additional path guaranteed to be empty, we show that sequential scanning, bright-port recycling, and Zeno-type interrogation all become particular feasible strategies within this same optimization, rather than separate benchmarks to compare against. This unified formulation identifies the optimal interaction-free localization strategy for any given set of resources, opening a route toward loss-resilient quantum imaging protocols for the study of fragile, absorption-sensitive samples.
\end{abstract}

\maketitle

\section{Introduction}

One of the most counterintuitive aspects of quantum mechanics is the possibility of obtaining information about a system without any direct interaction taking place. This idea was formalized in the Elitzur--Vaidman (EV) protocol for \emph{interaction-free measurement} (IFM), in the setting of detecting an ultra-sensitive absorbing object without triggering it~\cite{elitzur1993quantum}. The prototypical realization is a two-path interferometer built from \(50/50\) beam splitters and aligned so that, when no object is present, the photon always exits through the ``bright'' port $b$. If an absorber is placed in one of the arms, the interference is broken: the photon is absorbed with probability \(1/2\), exits the bright port with probability \(1/4\) (inconclusive event) or exits the dark port with probability \(1/4\). The dark-port event is the interaction-free success: it certifies the absorber's presence while the photon has not been absorbed. The ratio of interaction-free successes to all conclusive events gives the standard EV efficiency \(\eta_2=1/3\)~\cite{kwiat1995interaction}. This elementary two-path calculation contains the basic logic of conventional single-pass IFM.

Beyond its foundational role in demonstrating that information can be obtained without a direct interaction between the probe and the object of interest, interaction-free measurement has become a powerful tool for quantum sensing, imaging, and information processing. This idea has shaped low-damage imaging and sensing techniques for the study of fragile or photosensitive samples~\cite{White1998,OAMmultipixel,Cohen25}, interaction-free imaging with undetected photons~\cite{Yang2023}, and quantum-interrogation protocols based on repeated weak interaction and Zeno stabilization~\cite{kwiat99,Ma2014,Peise2015InteractionFree}. It has also led to counterfactual information-processing protocols, including counterfactual computation~\cite{mitchison2001counterfactual,Hosten2006} and counterfactual communication~\cite{salih2013protocol,Calafell:19,Wei-wei}. These developments broadened the platforms, improved the efficiency, and sharpened the notion of counterfactuality. Yet they largely preserve the same logical form as the original dark-port event: a successful no-absorption branch certifies that an object, obstruction, or channel condition is present.

A crucial enhancement of the original EV protocol invokes the quantum Zeno effect, using repeated weak interactions to suppress absorption and boost the IFM efficiency toward unity~\cite{kwiat1995interaction,kwiat99}. This high-efficiency scheme assumes a known empty reference arm to improve detection in the interrogated arm. Early implementations reached efficiencies of \(50\%\) and \(73\%\)~\cite{kwiat1995interaction,kwiat99}, and later realizations extended the same principle to integrated photonic devices and ultracold atoms~\cite{Ma2014,Peise2015InteractionFree}.

Beyond efficiency improvements, the foundational resources underlying quantum advantage in IFM have also been investigated. Contextuality (the inability of any noncontextual model to reproduce certain quantum statistics \cite{Budroni2022}) has been identified as the resource powering the quantum-over-classical advantage in the EV task: the interrogation efficiency exceeds the bound achievable by any generalized noncontextual model for asymmetric beamsplitter ratios ($\eta_2\rightarrow1/2$ when $T/R \rightarrow 0$), and the standard $\eta_2 = 1/3$ value corresponds exactly to the noncontextual ceiling~\cite{Wagner2024,Giordani2023}.

While IFM protocols typically address the binary question of \textit{whether} an absorber is present, a natural and practically important refinement is to determine \textit{where} it is located. In the Zeno variations of IFM there is no question about the location of the absorber. Rather, it is determined to be present or not at a known location.  Existing interaction-free localization schemes have approached the question about location by retaining the two-path module and arranging many such modules in sequence. Nakamura \emph{et al.} demonstrated this idea with a serial array of add-drop ring resonators, each functioning as a Zeno-like EV stage~\cite{Nakamura2023}. A simpler sequential EV version tests the \(d\) candidate paths one after another against a known reference path. If the tested path is empty, the photon exits the bright port deterministically and may be routed to the next stage; if the dark port clicks at stage \(j\), the absorber is identified as occupying path \(j\). Such schemes are natural optical baselines: they localize by sequential scanning. A separate line of work has extended IFM to the simultaneous detection of multiple objects: Filatov and Auzinsh proposed a cascaded EV scheme for this purpose~\cite{Filatov2024}. A modified version of their scheme was recently proposed and demonstrated experimentally on a programmable integrated photonic processor~\cite{Franco2026}. Across these approaches the no-absorption branch is read only as a binary witness of presence, even when many such witnesses are arranged in sequence to localize by scanning.
These schemes localize the absorber by unfolding the \(d\)-path problem into \(d\) sequential two-path decisions: each round commits one photon pass to a single candidate path, and the location is revealed only after as many as \(d\) such passes. We replace this temporal cascade with a single spatial structure: one \(d\)-path interferometer that interrogates every candidate location within the same photon pass, rather than one location at a time. By \emph{coherent interrogation}, we mean that the photon is prepared in a superposition across all candidate paths, the corresponding path amplitudes encounter the absorber region within the same use, and they are recombined before any which-path measurement is made.

\begin{figure}[t]
  \centering
  \includegraphics[width=\columnwidth]{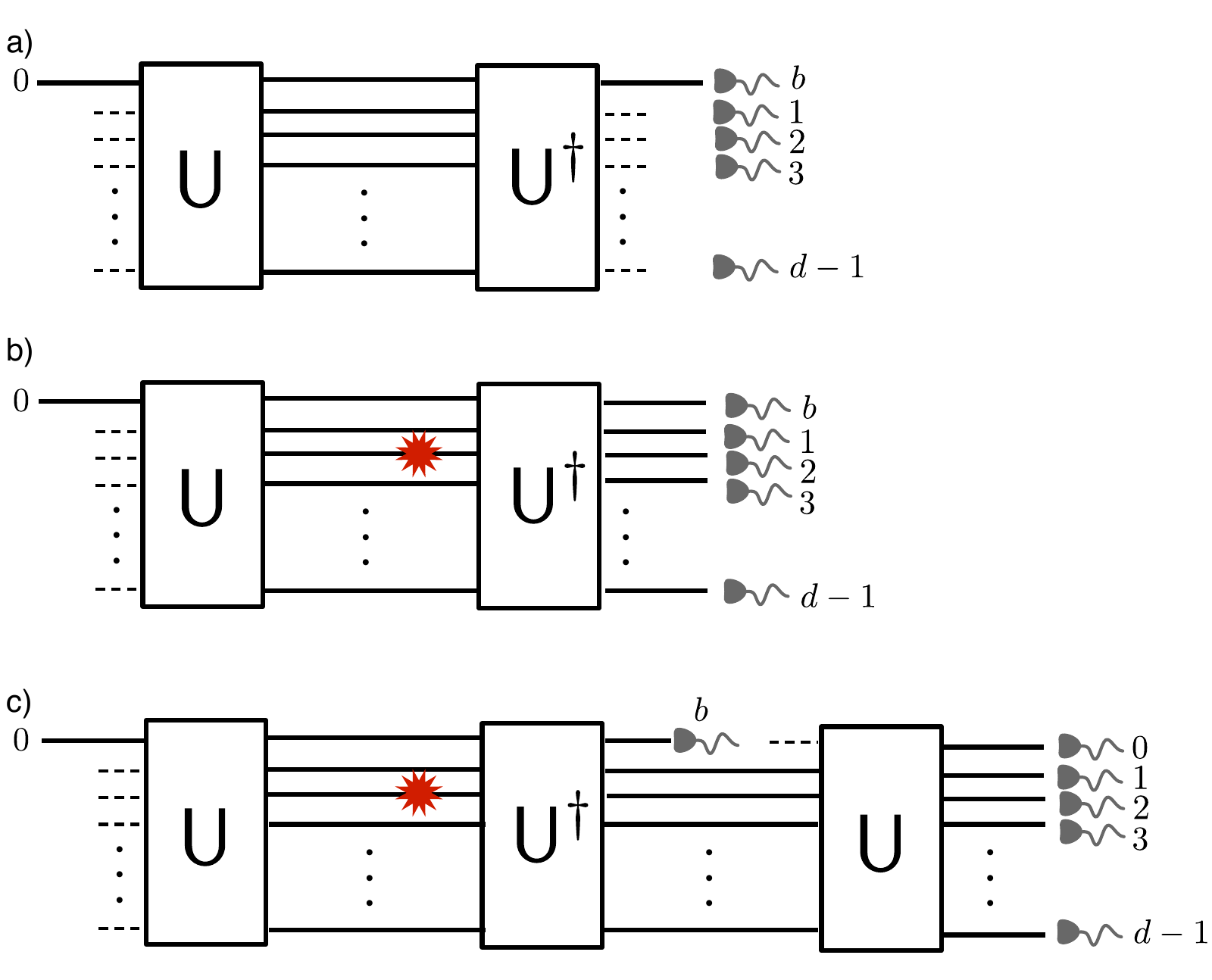}
  \caption{\label{fig:Int}
  High-dimensional interaction-free localization.
  (a) In the absence of an absorber, the balanced multipath interferometer sends the photon to the bright port \(b\).
  (b) An absorber in path \(a\) breaks the destructive interference. Conditional on no absorption, a dark-port click certifies the absorber's presence.
  (c) The dark-port output is a path-dependent state in the \((d-1)\)-mode dark subspace. A final balanced multiport beam splitter realizes the optimal readout and identifies the occupied path with probability \((d-1)/d\).
  }
\end{figure}

In this paper, we propose alternative strategies based on the coherent interrogation of multiple locations. Furthermore, we cast interaction-free localization as an operational task with a hard constraint: no location may be reported in the absence of an absorber. For a single perfect absorber, hidden uniformly among $d$ paths, we prove that the exact optimal success probability is attained by a multipath interferometer built with three balanced multiport beamsplitters, as shown in Fig. \ref{fig:Int}. The first beamsplitter prepares the probe state in a balanced superposition of the $d$ paths, while the second ensures the hard constraint by sending the photon to the bright port when there is no absorber. Conditioned on the informative dark-port branch, the absorber locations generate a regular-simplex ensemble of quantum states, and the third beamsplitter implements the optimal square-root measurement that discriminates among them. With this approach, the successful no-absorption branch becomes a spatially structured optical state, superseding the information carrier used in the EV-scheme and sequential strategies based on it. This becomes clear when we address multiple-absorber localization~\cite{Filatov2024,Franco2026}. For \(k\) absorbers among \(d\) paths, the no-absorption branch encodes the absorber subset collectively in one quantum state, rather than revealing occupied paths one by one. The subset-labelled dark-port states have closed-form overlaps, giving a structured minimum-error discrimination problem and an exact one-shot localization law for the balanced multipath architecture~\cite{Bergou10022010}.

We then lift the task to finite-round interaction-free localization using the quantum comb formalism~\cite{Chiribella2008,Chiribella2009}. The resulting adaptive comb-tester benchmark optimizes over all multi-pass optical testers in the chosen process model. The solved instances reveal strict adaptive advantage beyond the exact single-pass ceiling.

To compare the available strategies within a common operational model, we add one path that is promised to remain empty. Sequential, recycling, Zeno-type, and coherent multipath protocols then become feasible testers of the same reference-assisted process. For a single ideal absorber hidden uniformly in two candidate paths, the unassisted one-use optimum is $1/8$. A second unassisted use raises the exact value to $27/128$, while the finite EV scan, which uses the promised empty path, reaches $1/4$. Optimizing over all two-use causal strategies with that same reference path gives the strictly larger value $25/72$. The optimum is attained by a coherent two-round strategy: the probe is prepared with path weights $(p_r,p_0,p_1)=(1/2,1/4,1/4)$, the branch surviving the first encounter is mapped by an isometry to a joint path-memory state, and the resulting path component undergoes the second encounter before the final Helstrom measurement. Thus the second interrogation is causally generated from the first surviving output. The strategy is genuinely adaptive in the quantum-comb sense, although no intermediate measurement or classical feedforward is required.

Finally, we address the interaction-free localization problem under non-ideal conditions, such as optical losses or other imperfections. In particular, we show that the use of multipath interferometers with recycling of the bright port is more robust than a sequential Zeno scan under optical losses. Adaptive gains are numerically found for a number of scenarios.

\section{Balanced multiport interaction-free localization}
\label{sec:balanced_multiport}

We now describe the interferometric circuit that turns the dark port from a witness of presence into a carrier of position. We consider optical interferometers for convenience, as translation of our results to other platforms is straightforward. The construction is the direct multipath analogue of the EV interferometer, as shown in Fig.~\ref{fig:Int}. It consists of three balanced multiport beam splitters described by a dephased complex Hadamard unitary
\begin{equation}
\begin{aligned}
  \oper{U}
  &=
  \frac{1}{\sqrt d}
  \sum_{j,k=0}^{d-1} u_{jk}\ket{j}\bra{k},
  \qquad |u_{jk}|=1,\\
  u_{j0}&=u_{0k}=1 .
\end{aligned}
\label{eq:balanced_multiport}
\end{equation}
Fourier and Hadamard transformations~\cite{carine20} are particular examples of Eq. \eqref{eq:balanced_multiport}. An input photon in mode \(\ket{0}\) is sent by the first multiport into an equal superposition over the \(d\) paths. If no absorber is present, the second multiport recombines the state back to mode \(0\). Thus a photon in the empty interferometer exits through the bright port \(b\) with certainty and never produces a dark-port click.

\begin{figure}[t]
  \centering
  \includegraphics[width=\columnwidth]{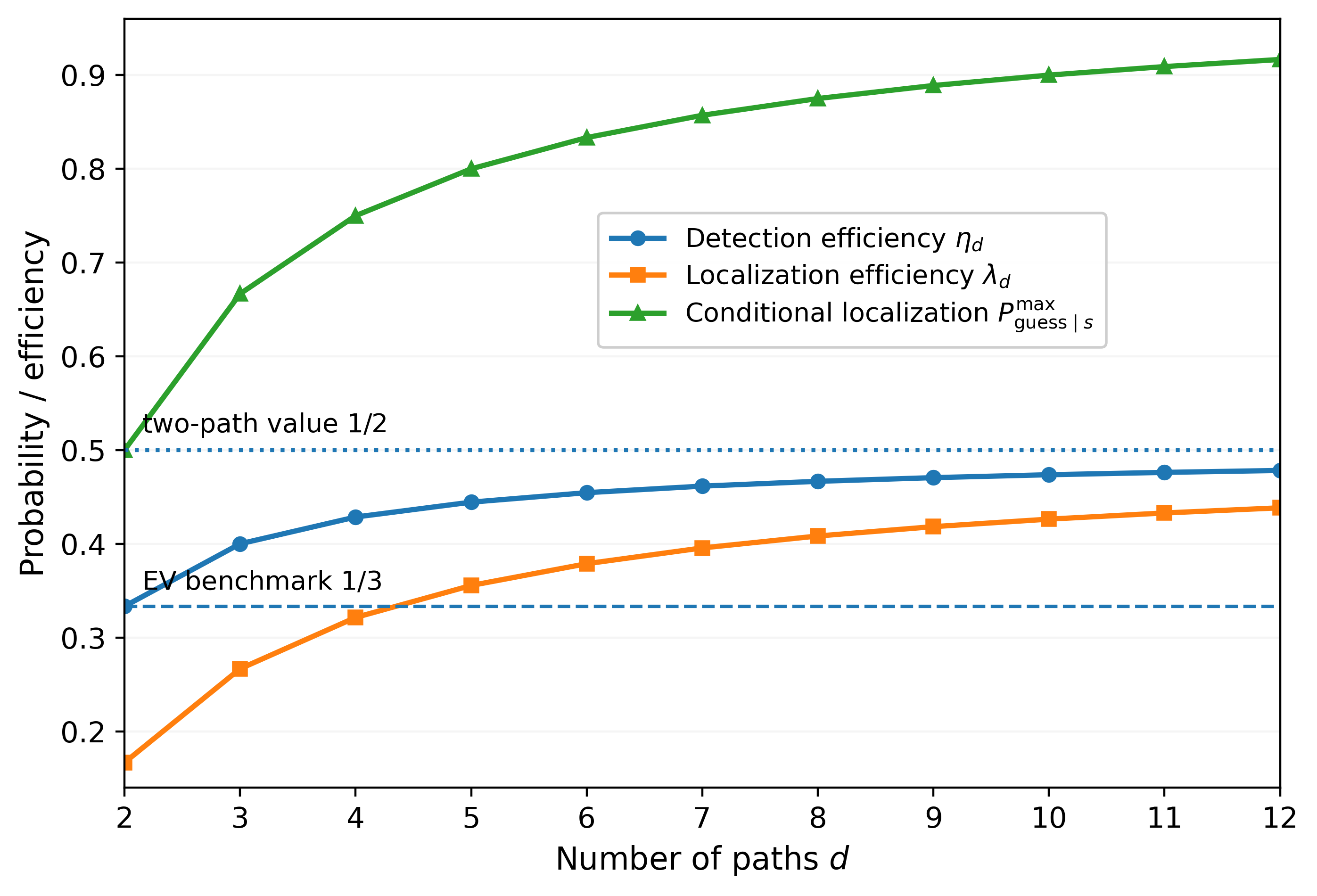}
  \caption{\label{fig:eta_guess}
  Dimension scaling of the balanced multiport protocol.
  The detection efficiency \(\eta_d\), localization efficiency \(\lambda_d\), and conditional dark-branch guessing probability \(P_{\mathrm{guess}\mid s}^{\max}\) are plotted against the number of paths \(d\).
  The dashed line marks the EV benchmark \(1/3\); the dotted line marks the two-path conditional value \(1/2\), which is the limit of the detection efficiency in the unbalanced EV scheme when $R\to 0$.
  Detection exceeds the EV value for \(d\ge3\), while the dark branch carries nontrivial location information for every genuinely multipath interferometer.
  }
\end{figure}

We now consider an ideal absorber in path \(a\). Let us define operators \(\oper{A}=\ket{a}\bra{a}\) and \(\oper{N}=\oper{I}-\oper{A}\) to represent absorption and non-absorption, respectively. Conditioned on survival, the photon exits the second beam splitter in the state
\begin{equation}
  \ket{\psi_e}
  =
  \oper{U}^{\dagger}\oper{N}\oper{U}\ket{0}
  =
  \ket{0}
  -
  \frac{1}{d}\sum_{n=0}^{d-1}u^*_{an}\ket{n}.
  \label{eq:exit_state}
  \end{equation}
  The \(n=0\) term in \(\ket{\psi_e}\) is the bright-port amplitude, and the remaining modes form the dark component. Using \eqref{eq:exit_state}, the outcome probabilities are
  \begin{equation}
  \begin{aligned}
  P_{d*}&=\frac1d,\qquad
  P_{db}=\left(\frac{d-1}{d}\right)^2,\qquad
  P_{ds}=\frac{d-1}{d^2},
\end{aligned}
\label{eq:single_absorber_probabilities}
\end{equation}
where \(P_{d*}\), \(P_{db}\), and \(P_{ds}\) are the absorption, bright-port, and dark-port probabilities.  The no-absorption exit probability is \(P_{de}=1-P_{d*}=(d-1)/d\).
We can then find the efficiency as
 \begin{equation}
  \eta_d
  :=
  \frac{P_{ds}}{1-P_{db}}
  =
  \frac{P_{ds}}{P_{ds}+P_{d*}}
  =
  \frac{d-1}{2d-1}.
\label{eq:single_absorber_efficiency}
\end{equation}
  The efficiency \(\eta_d\) equals the EV value \(1/3\) at \(d=2\), is strictly larger for \(d\ge3\), and tends to \(1/2\), as shown in Fig.~\ref{fig:eta_guess}. This is the same $1/2$ ceiling approached by the tunable two-path EV scheme \cite{elitzur1993quantum,Wagner2024}, which has also been recently pointed out in Ref.~\cite{Franco2026}. Bright-port events may be recycled as in EV: in the empty interferometer the photon never reaches a dark port, while with an absorber repeated recycling eventually leads either to absorption or to a dark-port event~\cite{kwiat1995interaction}.

The two-path EV interferometer has one dark event. The \(d\)-path interferometer has a \((d-1)\)-dimensional dark subspace, which is what makes localization possible. Conditioning on no absorption and no bright-port click gives a location-dependent output state
\begin{equation}
  \ket{\phi_{sa}}
  =
  -\frac{1}{\sqrt{d-1}}
  \sum_{n=1}^{d-1}u^*_{an}\ket{n}.
  \label{eq:dark_port_simplex_state}
  \end{equation}
These states have overlap
  $\left|\langle\phi_{sa}|\phi_{sa'}\rangle\right|
  =
  1/(d-1)$
  for $a\neq a'$, which follows from unitarity of \(\oper{U}\) and the dephased convention.
Thus, the \(d\) possible absorber locations prepare an equiprobable regular-simplex ensemble in the \((d-1)\)-dimensional dark subspace. The maximum probability of determining which state is present is known and given by
 \begin{equation}
 P_{\text{guess}\mid s}^{\max}
  =
  \frac{d-1}{d},
\label{eq:dark_port_simplex_probability}
\end{equation}
 where the optimal minimum-error readout is obtained by the symmetric square-root measurement~\cite{Bergou10022010}. For \(d=2\), \(P_{\text{guess}\mid s}^{\max} = 1/2\), so the dark branch contains no nontrivial location information. For every \(d\ge3\), the dark branch carries position information.

Multiport interferometers with auxiliary modes can implement general POVMs~\cite{Martinez2023Certification,martinez2025,yasir2025compactifyinglinearopticalunitaries}. Here the symmetry of the dark-port ensemble makes the optimal readout another balanced multiport splitter.  Acting with a final copy of \(\oper{U}\) on the dark state (see Fig. \ref{fig:Int}) gives
\begin{equation}
  \oper{U}\ket{\phi_{sa}}
  =
  -\sqrt{\frac{d-1}{d}}\ket{a}
  +
  \frac{1}{\sqrt{d(d-1)}}\sum_{n\neq a}\ket{n}.
  \label{eq:multiport_readout_state}
  \end{equation}
  A click in output \(a\) identifies the occupied path with probability \((d-1)/d\), while \(P_{x\neq a}=1/[d(d-1)]\) for each wrong output, reproducing the square-root measurement value.
  The probability of successfully locating the absorber is then
  \begin{equation}
  P_{\rm loc}^{\rm mp}(d)
  :=
  P_{ds}P_{\text{guess}\mid s}^{\max}
  =
  \frac{(d-1)^2}{d^3}.
  \label{eq:multiport_readout_probability}
  \end{equation}

We quantify localization performance by the localization efficiency $\lambda_d$, the probability of a correct interaction-free location report per conclusive event, in direct analogy with the detection efficiency $\eta_d$:
\begin{equation}
  \lambda_d
  :=
  \frac{P_{ds}P_{\text{guess}\mid s}^{\max}}{1-P_{db}}
  =
  \frac{(d-1)^2}{d(2d-1)}.
\label{eq:multiport_readout_efficiency}
\end{equation}
Its dimension scaling, together with $P_{\mathrm{guess}\mid s}^{\max}$, is shown in Fig.~\ref{fig:eta_guess}. Like $\eta_d$, the efficiency $\lambda_d$ grows with $d$ and approaches $1/2$. The natural baseline is a sequential EV scan, which localizes by testing one path at a time: with a symmetric $50/50$ beam splitter its per-conclusive-event efficiency is the familiar $1/3$, but biasing the beam splitter raises it toward $1/2$ ~\cite{elitzur1993quantum,Wagner2024,Franco2026}. The multipath protocol therefore does not beat an optimized sequential scan on this lossless ratio. Its advantage is structural rather than ratio-based: it localizes in a single coherent pass over all $d$ paths, rather than a depth-$d$ cascade of two-path tests.

Below we prove that the multipath protocol is optimal in the ``one-shot'' scenario, where a single interrogation time slot is allowed. Moreover, the success probability $P_{\rm loc}^{\rm mp}(d)$ cannot be improved by the use of ancillary systems nor more general measurements satisfying the interaction-free constraint. We then show that a single-pass protocol becomes a decisive operational gain under loss, and that it extends to the collective readout of several absorbers and to adaptive strategies. {Sequential scanning will be revisited in a general framework for adaptive strategies developed in section \ref{sec:adaptive_ifm}.}

\section{Exact one-shot optimality}
\label{sec:one_shot_optimality}

The balanced multiport achieves one-shot localization with success probability \(P_{\rm loc}^{\rm mp}(d)=(d-1)^2/d^3\). We now show that this is the one-shot law, i.e., it is the optimal success probability while being confined to a single pass.  The optimization allows an arbitrary probe state, an arbitrary finite-dimensional ancilla, and an arbitrary final measurement.

Let \(\mathcal H_S=\mathrm{span}\{\ket{0},\ldots,\ket{d-1}\}\) be the path space, and \(\mathcal H_A\) the space of an arbitrary finite-dimensional ancilla. Let a single ideal absorber be hidden uniformly in path \(a\). The no-absorption branch operator is \(\oper{N}_a=\oper{I}-\ket{a}\bra{a}\). A general strategy prepares \(\rho_{SA}\), applies \(\oper{N}_a\otimes\oper{I}_A\), and measures with conclusive effects \(M_a\) and an inconclusive effect \(M_?\). Its one-shot localization probability is

\begin{align}
  P_{\rm loc}
  &=
  \frac1d\sum_{a=0}^{d-1}
  \operatorname{Tr}\!\left[
  (\oper{N}_a\otimes\oper{I}_A)\rho_{SA}
  (\oper{N}_a\otimes\oper{I}_A)M_a
  \right],
  \label{eq:general_one_shot_task}\\
  \operatorname{Tr}(\rho_{SA}M_a)&=0
  \qquad \forall a .
  \label{eq:empty_interferometer_constraint}
\end{align}
The second line is the empty-interferometer condition. It says that no absorber label can be produced by a state that would also appear when no absorber is present.

\begin{theorem}[Exact one-shot interaction-free localization]
\label{thm:one_shot_optimality}
For a single ideal absorber hidden uniformly among \(d\) paths, the optimal one-shot localization probability is
\begin{equation}
  P_{\rm loc}^{\star}(d)=\frac{(d-1)^2}{d^3}.
  \label{eq:one_shot_optimal_value}
\end{equation}
Ancillary systems do not improve this value, and equality is attained by the balanced multipath protocol described above.
\end{theorem}

The proof, provided in detail in Appendix \ref{app:one_shot_proof}, reduces the constraint \eqref{eq:empty_interferometer_constraint} to the geometry of the informative no-absorption branch. Here we provide a brief sketch. By purification and linearity, it is enough to consider a pure input state \(\ket{\Psi}\). The empty-interferometer condition and positivity of \(M_a\) imply \(M_a\ket{\Psi}=0\) for every conclusive outcome. Thus, the component parallel to the empty-interferometer state is unavailable for localization. All useful information lies in the absorber-induced component orthogonal to \(\ket{\Psi}\). With \(Q=\oper{I}-\ket{\Psi}\bra{\Psi}\), define the informative vectors \(\ket{w_a}=Q(\oper{N}_a\otimes\oper{I}_A)\ket{\Psi}\). The original task is thereby reduced to minimum-error discrimination of the subnormalized pure states \(d^{-1}\ket{w_a}\bra{w_a}\).

The use of ancilla cannot enlarge this geometry. Writing \(\ket{\Psi}=\sum_i\ket{i}\ket{\eta_i}\), with \(p_i=\langle\eta_i|\eta_i\rangle\), one finds that the Gram matrix of the informative vectors depends only on the path weights \(p_i\). The same Gram matrix is produced by the signal-only state \(\ket{\psi_p}=\sum_i\sqrt{p_i}\ket{i}\). Since the optimal discrimination probability of pure states is fixed by their Gram matrix, every ancilla-assisted strategy has an ancilla-free counterpart with the same performance.

The remaining problem is therefore a path-weight optimization. For a signal-only probe with weights \(p=(p_0,\ldots,p_{d-1})\), the informative ensemble obeys the dual bound
\begin{equation}
  P_{\rm loc}(p)
  \le
  \frac{d-1}{d^2}\left(1-\sum_{i=0}^{d-1}p_i^2\right).
  \label{eq:path_weight_bound}
\end{equation}
The right-hand side is maximized uniquely by the uniform distribution \(p_i=1/d\). This gives \(P_{\rm loc}(p)\le(d-1)^2/d^3\). The balanced probe saturates the bound: its informative states are precisely the regular-simplex dark-port states constructed above, and the square-root measurement attains the value in Eq.~\eqref{eq:one_shot_optimal_value}. This proves Theorem~\ref{thm:one_shot_optimality}.

The theorem explains the boundary between two-path and multipath IFM. In two paths, the interaction-free branch has no nontrivial location space. With \(d\ge3\), the orthogonal no-absorption subspace can support absorber-labelled simplex states. The balanced multiport is the equality case of the one-shot bound: it creates the maximum possible absorber-induced distinguishability allowed by the interaction-free constraint.

\section{Multiple absorbers: collective dark-port states}
\label{sec:multiple_absorbers}

Having fixed the one-shot limit for a single absorber, we now ask what the same multipath architecture records when several absorbers are present. Whereas a sequential two-path scan terminates at the first successful dark click without scanning all the paths or needs an ancillary system with a dimension that increases exponentially with the number of locations interrogated, the multipath protocol leaves on the no-absorption branch a single state labelled by the entire occupied subset.

Let \(\mathcal A\subset\{0,\ldots,d-1\}\) be the occupied set, with \(|\mathcal A|=k\). We write the absorption operator as \(\oper{A}_{\mathcal A}=\sum_{a\in\mathcal A}\ket{a}\bra{a}\) and \(\oper{N}_{\mathcal A}=\oper{I}-\oper{A}_{\mathcal A}\) for non-absorption. For the balanced multiport protocol, the output quantum state is
\begin{equation}
  \ket{\psi_e^{\mathcal A}} = \oper{U}^{\dagger}\oper{N}_{\mathcal A}\oper{U}\ket{0} = \ket{0} - \frac1d \sum_{a\in\mathcal A}
  \sum_{n=0}^{d-1}u^*_{an}\ket{n}.
  \label{eq:k_absorber_exit_state}
  \end{equation}
  The outcome probabilities are
  \begin{equation}
  P_{d*}^{(k)} = \frac{k}{d},\;
  P_{db}^{(k)} = \left(\frac{d-k}{d}\right)^2,\;
  P_{ds}^{(k)} = \frac{k(d-k)}{d^2},
    \label{eq:k_absorber_pb}
  \end{equation}
  which give detection efficiency
  \begin{equation}
  \eta_{d,k}
  =
  \frac{d-k}{2d-k}.
  \label{eq:k_absorber_ps_eta}
 \end{equation}
Thus, the dark branch exists precisely when \(0<k<d\). Its probability depends only on the sizes of the occupied and unoccupied path sets, not on the particular subset \(\mathcal A\).

Conditioning on no absorption and no bright-port click gives the dark-branch output state
\begin{equation}
  \ket{\phi_{s\mathcal A}}
  =
  -\frac{1}{\sqrt{k(d-k)}}
  \sum_{a\in\mathcal A}
  \sum_{n=1}^{d-1}u^*_{an}\ket{n}.
  \label{eq:k_absorber_dark_state}
  \end{equation}
These states have overlap given by
  \begin{equation}
  \left|\langle\phi_{s\mathcal A'}|\phi_{s\mathcal A}\rangle\right|
  =
  \left|
  \frac{d|\mathcal A\cap\mathcal A'|-k^2}
       {k(d-k)}
  \right|,
  \label{eq:k_absorber_overlap}
\end{equation}
which depends only on the size of the intersection \(x=|\mathcal A\cap\mathcal A'|\). Some subset states are therefore orthogonal while others coincide. In a \(d=4\) interferometer with \(k=2\), for instance, any two subsets that share a single path are orthogonal, so a single dark-branch photon distinguishes them perfectly. Only complementary subsets, which share no path, collapse onto the same state. The no-absorption branch thus stores relational information about the whole configuration, with the geometry fixed entirely by the subset intersections. This is the qualitative difference from a sequential scan: in the multipath protocol, the photon is distributed over all \(d\) paths before the no-absorption branch is selected, and the resulting state \(\ket{\phi_{s\mathcal A}}\) carries information about the whole subset \(\mathcal A\).

Let \(\rho_{\mathcal A}=\ket{\phi_{s\mathcal A}}\bra{\phi_{s\mathcal A}}\). For prior probabilities \(p_{\mathcal A}\) and relevance weights \(c_{\mathcal A}\), the optimal readout for the balanced architecture is the finite minimum-error discrimination problem
\begin{align}
  G_k^{\rm opt}
  =
  \max_{\{\Pi_{\mathcal A}\}}
  &\sum_{\mathcal A}
  c_{\mathcal A}p_{\mathcal A}
  \operatorname{Tr}\!\left(\Pi_{\mathcal A}\rho_{\mathcal A}\right),
  \label{eq:k_absorber_sdp}\\
  \text{subject to}\quad
  &\sum_{\mathcal A}\Pi_{\mathcal A}=\oper{I},
  \qquad
  \Pi_{\mathcal A}\succeq0\quad(\forall\,\mathcal A).
  \nonumber
\end{align}
For uniform priors and equal relevance weights, this readout also has a closed analytic value. The subset states form a symmetric tight frame in the \((d-1)\)-dimensional dark subspace, and the square-root measurement is optimal. Hence
\begin{equation}
  G_k^{\rm opt}
  =
  \frac{d-1}{\binom dk},
  \qquad
  P_{\rm loc}^{(d,k)}
  =
  \frac{k(d-k)}{d^2}\,
  \frac{d-1}{\binom dk}.
  \label{eq:k_absorber_analytic_value}
\end{equation}
For \(k=1\) this reduces to the single-absorber value. For \(d=4,k=2\), it gives \(G_2^{\rm opt}=1/2\): the six subsets collapse into three orthogonal complementary pairs. The proof and representative values are given in Table~\ref{tab:app_exact_small_values} and in Appendix \ref{app:k_absorbers}.

The physical distinction from the single-absorber case is modest but important. The same balanced preparation, absorber region, and recombination stage produce the dark subset states. For one absorber, the symmetry of the simplex makes a final balanced multiport the optimal readout. For several absorbers, the natural object is the subset-state ensemble itself, and the optimal readout is generally the finite POVM in Eq.~\eqref{eq:k_absorber_sdp}, where special symmetric cases reduce to simple analytic measurements.

Multipath IFM therefore changes the information carrier. A successful no-absorption branch need not encode a single encountered location, it can encode a whole absorber configuration. The next section turns from one-shot multiport architectures to finite-round adaptive strategies, where the interaction-free localization task is optimized by the comb-tester benchmark.

\section{Adaptive advantage in interaction-free localization}
\label{sec:adaptive_ifm}

The one-shot theorem sets the boundary for a single encounter with the absorber region. Adaptive IFM begins when the surviving branch is not discarded, but kept as a physical resource. The surviving component of the probe, which previously led to either a bright or a dark outcome, now may be stored, fed back into the interaction region, interfered with additional modes, or measured. Later inputs may depend on earlier outcomes, and the final report is accepted only if no absorption has occurred. The object stays fixed throughout, while the strategy now carries memory across uses.

Optimizing over \emph{all} finite-round strategies, including those that store the surviving photon, feed it back, and condition later operations on earlier outcomes, calls for a description of the apparatus as a whole rather than channel by channel. The quantum-comb formalism supplies exactly this: it represents an entire multi-time network as a single object (a higher-order map) with open ``slots'', generalizing the Choi representation of a channel to a process that acts at several times with memory in between~\cite{Chiribella2008,Chiribella2009}. Its value here is that it cleanly separates the fixed object from the strategy: the network with open slots, called the \emph{tester}, is what we optimize over, while the process inserted into the slots is held fixed. Here the inserted process is the non-absorbing or surviving branch of the hidden absorber configuration. Let \(j\) label that configuration: for one absorber, \(j\) is a path label; for several absorbers, \(j\) labels a subset.

For one use, the object is described by a completely positive trace-nonincreasing operation
\begin{equation}
  \mathcal S_j(\rho)
  =
  \sum_{\mu}K_{j\mu}\rho K_{j\mu}^{\dagger},
  \qquad
  \sum_{\mu}K_{j\mu}^{\dagger}K_{j\mu}\le \oper I .
  \label{eq:no_click_process}
\end{equation}
This operation may include attenuation, phase shifts, mode mixing, scattering into auxiliary modes, detector internal degrees of freedom, and propagation loss, as we will discuss in Section \ref{sec:implementation}. The perfect absorber used in the one-shot theorem is the special case
\(
\mathcal S_j(\rho)=\oper{N_j}\rho\oper{N_j}
\), with \(\oper{N_j}=\oper I-\oper{A_j}\), where \(\oper{A_j}\)
projects onto the occupied path or occupied subspace.

With the standard Choi convention, let \(\sigma_j=J(\mathcal S_j)\) be the Choi operator for the single-pass process (see details in Appendix \ref{app:Choi_rep}). For \(K\) independent uses of the same hidden object, the surviving process is
\begin{equation}
  \Sigma_j=\sigma_j^{\otimes K}.
  \label{eq:k_pass_survival_process}
\end{equation}
The tensor product describes the repeated object. The strategy is the causal tester connecting the slots, and it may contain internal memory, intermediate interferometers, feedforward, and a final measurement.

Let \(\mathsf{Test}_K\) denote the cone of \(K\)-round testers obeying the causal normalization constraints, that is, allowing information to flow from earlier slots to later slots, but never from later slots to earlier ones (details in Appendix \ref{app:adaptive_strategies}). A conclusive report \(j\) is represented by a positive tester element \(T_j\). If the true configuration is \(j\), the probability of a correct localization report is \(\operatorname{Tr}(T_j\Sigma_j)\). Let \(\Sigma_{\varnothing}\) denote the empty-interferometer process. The interaction-free tester cone is
\begin{equation}
  \mathsf{Test}_K^{\rm IF}(\Sigma_{\varnothing})
  =
  \left\{
  \{T_j\}\in\mathsf{Test}_K:
  \sum_{j=1}^{N}
  \operatorname{Tr}\!\left(T_j\Sigma_{\varnothing}\right)=0
  \right\} .
  \label{eq:if_tester_cone}
\end{equation}
Since all terms are nonnegative, this is equivalent to \(\operatorname{Tr}(T_j\Sigma_{\varnothing})=0\) for every conclusive branch. The adaptive interaction-free localization value is
\begin{equation}
  P_{\rm IF}^{(K)}
  =
  \max_{\{T_j\}\in\mathsf{Test}_K^{\rm IF}(\Sigma_{\varnothing})}
  \frac1N
  \sum_{j=1}^{N}
  \operatorname{Tr}\!\left(T_j\Sigma_j\right),
  \label{eq:adaptive_ifm_value}
\end{equation}
where \(N\) is the number of possible absorber configurations.

Because Eq.~\eqref{eq:adaptive_ifm_value} is written as an
optimization over the abstract tester cone, it is not immediately clear
that its value is physically attainable, nor that no physical strategy
can exceed it. The comb realization theorem (see Eq.~(67) and Theorem~11 in Ref.~\cite{Chiribella2009}) closes both gaps: every
physical adaptive strategy corresponds to a feasible tester, and every
feasible tester can be realized physically, so the two optimizations
coincide. This yields the following benchmark.

\begin{theorem}[Adaptive IFM benchmark]
\label{thm:adaptive_ifm_benchmark}
For fixed surviving processes \(\{\Sigma_j\}_{j=1}^{N}\), empty-interferometer process \(\Sigma_{\varnothing}\), and with the tester constraints and the empty-device condition in Eq.~\eqref{eq:if_tester_cone}, Eq.~\eqref{eq:adaptive_ifm_value} gives the maximum probability of identifying the absorber configuration without absorption over all \(K\)-use adaptive quantum strategies compatible with the process model.
\end{theorem}

The benchmark is exact in both directions. Every physical adaptive
experiment induces feasible tester elements whose pairing with the
process operators reproduces its outcome probabilities, and every
feasible tester admits a sequential realization by a preparation,
memory-carrying intermediate operations, and a final measurement. At
\(K=1\), the construction reduces to the one-shot discrimination problem.
For \(K>1\), it gives the finite-round interaction-free localization
problem itself.

This formulation places the standard interferometric constructions inside one causal class, so that strategies usually treated as distinct protocols appear instead as different testers for the same process. A sequential scan routes the probe through candidate locations in order \cite{Nakamura2023}; a recycling strategy sends an inconclusive bright-port branch into a later use \cite{elitzur1993quantum}; a Zeno-type protocol distributes weak interactions across many slots \cite{kwiat1995interaction}; and a multipath adaptive circuit keeps candidate paths coherent while using earlier survival information to choose later wiring. The cascaded multi-object schemes of Filatov and Auzinsh, realized experimentally by Franco~\textit{et~al.}, fall in the sequential-recycling family, where the surviving photon is passed from one two-path stage to the next~\cite{Filatov2024,Franco2026}. It is important to remark that a sequential two-path scan requires one additional path that is promised not to contain the absorber. This extra ``free rail'' is a strategy resource rather than an additional absorber hypothesis. It splits the strategies in two classes: we call a strategy \emph{unassisted} when no additional path is promised to remain empty, or \emph{reference-assisted} when it includes a free rail. Once the available geometry is fixed, sequential, recycling, Zeno-type, and coherent multipath protocols are feasible testers within the corresponding class.

\begin{figure*}[t]
  \centering
  \includegraphics[width=0.8\textwidth]{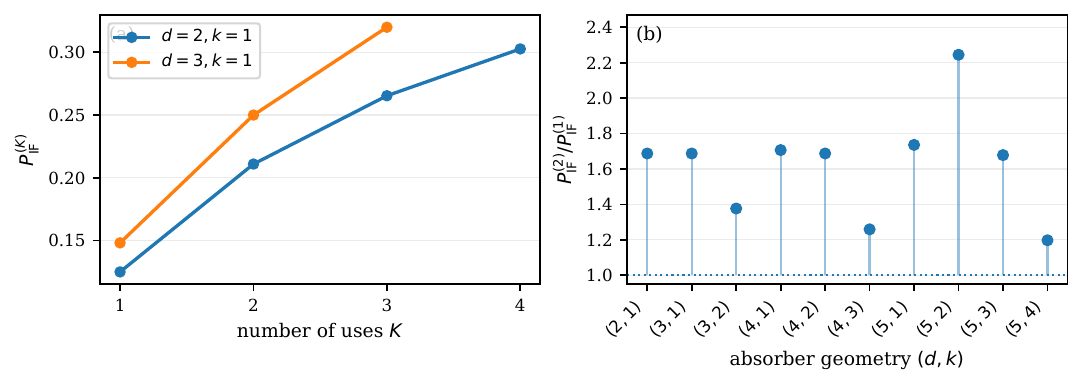}
  \caption{\label{fig:adaptive_advantage}
  Adaptive advantage over single-shot optimal strategy. (a) Optimal interaction-free localization probability \(P_{\rm IF}^{(K)}\) versus the number of uses \(K\) for one absorber in two and three paths.
  (b) Two-use adaptive gain \(P_{\rm IF}^{(2)}/P_{\rm IF}^{(1)}\) for all solved geometries with \(K=2\).
  The dotted line marks no adaptive advantage.
  The corresponding numerical values are listed in Table \ref{tab:app_adaptive_ideal_values}.}

\end{figure*}

Let us focus for now on unassisted strategies. Solving these tester optimizations shows that adaptivity strictly increases the interaction-free localization probability, with the gain growing in the number of uses \(K\) (Fig.~\ref{fig:adaptive_advantage}). For the setting of a single absorber in two paths, it improves from the one-shot value \(1/8\) to
\(0.21\) at two uses and \(0.30\) at four; in three paths it improves from
\(4/27\) to \(0.25\) at two uses and \(0.32\) at three. The effect persists
beyond a single absorber: for \((d,k)=(5,2)\), two uses more than double
the single-use value, a factor of \(2.24\). Numerical values for all solved
instances are collected in Table \ref{tab:app_adaptive_ideal_values} (see Appendix \ref{app:adaptive_tables_unassisted}).

These results separate the two resources the protocol draws on. The multi-path interferometer supplies a spatial resource, a dark subspace wide enough to carry location in a single pass, while adaptive IFM supplies a temporal one, a no-absorption history that can be stored, recombined, and measured jointly before the final report. Holding the absorber configuration fixed and varying only the tester, Fig.~\ref{fig:adaptive_advantage} isolates this temporal resource and shows that suitable temporal wiring raises the localization probability beyond the single-use optimum.

Now let us consider the minimal finite-round instance, which can be solved using strategies with different resources available. Consider one ideal absorber hidden uniformly in one of two candidate paths. We call the strategy \emph{unassisted} when no additional path is promised to remain empty.  Arbitrary ancillary systems and coherent memory are still allowed. The one-shot theorem gives \(P_{\rm IF}^{(1)}(2,1)=1/8\), where \(P_{\rm IF}^{(K)}(d,k)\) is the benchmark value in Eq. \eqref{eq:adaptive_ifm_value} for $d$-paths and $k$ absorbers.  A second use
raises this limit exactly.

\begin{proposition}[Exact unassisted two-use optimum]
\label{prop:exact_unassisted_two_use}
For one ideal absorber hidden uniformly in two candidate paths, with no
promised empty reference rail, the maximum lossless interaction-free
localization probability over all two-use causal strategies is
\begin{equation}
  P_{\rm IF}^{(2)}(2,1)
  =
  \frac{27}{128}.
  \label{eq:exact_unassisted_two_use}
\end{equation}
\end{proposition}

The proof is given in Appendix~\ref{app:two_use_exact}. The bound is
attained by a fixed coherent circuit comprising an asymmetric input
state, an intermediate path rotation, and a final measurement. Thus, no
intermediate measurement or feedforward is needed to attain the optimum,
even though Proposition~\ref{prop:exact_unassisted_two_use} optimizes over
the full two-use adaptive class. The exact gain over one use is
\begin{equation}
  \frac{P_{\rm IF}^{(2)}(2,1)}
       {P_{\rm IF}^{(1)}(2,1)}
  =
  \frac{27}{16}.
  \label{eq:exact_unassisted_two_use_gain}
\end{equation}

The standard finite two-path EV scan reaches a larger value. It introduces
one additional rail \(r\), promised not to contain the absorber, and tests
the two candidate paths successively against it.
Using beam splitters of reflectivity $R$, an EV stage localizes an absorber in the tested path with probability $R(1-R)\leq 1/4$, achieving its best performance with balanced beam splitters.

 Since
either candidate is tested within two uses,
\begin{equation}
  P_{\rm scan}^{(2)}(2,1)
  =
  \frac12\frac14+\frac12\frac14
  =
  \frac14
  >
  \frac{27}{128}.
  \label{eq:finite_scan_two_use}
\end{equation}

This inequality is not a comparison within one feasible set. The promised
rail is not a third absorber hypothesis; it is an additional strategy
resource. The scan therefore exceeds the unassisted optimum because it
addresses the same two-location task with an enlarged physical architecture.
The resource-matched question is whether sequential scanning is optimal
when every two-use causal strategy is supplied with that same rail. This
optimization also has a closed solution.

\begin{theorem}[Exact reference-assisted two-use optimum]
\label{thm:exact_reference_assisted_two_use}
For one ideal absorber hidden uniformly in two candidate paths, the
maximum lossless interaction-free localization probability over all
two-use causal strategies supplied with one promised empty reference rail
is
\begin{equation}
  P_{\rm IF,ref}^{(2)}(2,1)
  =
  \frac{25}{72}.
  \label{eq:exact_reference_assisted_two_use}
\end{equation}
The optimization allows arbitrary ancillary systems, coherent memory,
intermediate operations, and final measurements.
\end{theorem}

The proof is given in Appendix~\ref{app:two_use_exact}. Since
\begin{equation}
  \frac{25}{72}
  >
  \frac14,
  \label{eq:reference_assisted_beats_scan}
\end{equation}
the finite scan is strictly suboptimal within the reference-assisted
class. The full causal optimum exceeds the scan by the exact factor
\begin{equation}
  \frac{P_{\rm IF,ref}^{(2)}(2,1)}
       {P_{\rm scan}^{(2)}(2,1)}
  =
  \frac{25}{18}.
  \label{eq:reference_assisted_scan_gain}
\end{equation}
The reference rail makes the sequential scan possible, but sequential
interrogation does not use that rail optimally. Coherent causal processing
can retain and recombine the surviving amplitudes across the two uses and
thereby achieve a larger localization probability with the same absorber
hypotheses, the same number of uses, and the same promised rail.

The resulting exact hierarchy is
\begin{equation}
  \frac18
  <
  \frac{27}{128}
  <
  \frac14
  <
  \frac{25}{72}.
  \label{eq:exact_two_use_hierarchy}
\end{equation}
Its three strict inequalities isolate three distinct steps: a second
encounter with the absorber region, the addition of a promised empty rail,
and the coherent optimization of that rail.  The reference-assisted optimum is attained by a coherent two-round strategy that is genuinely adaptive in the quantum-comb sense. The probe is prepared with path weights $(p_r,p_0,p_1)=(1/2,1/4,1/4)$. Conditional on survival of the first encounter, a fixed isometry maps the path amplitudes to a joint path-memory state and thereby generates the coherent path state entering the second encounter. The final Helstrom measurement is performed only after the second surviving branch is obtained. Hence the second input is produced from the output of the first use, rather than prepared independently or selected as the next step of a fixed scan. The adaptation is coherent, so neither an intermediate measurement nor classical feedforward is required. Appendix~\ref{app:two_use_exact} gives the exact isometry and final POVM.

\begin{figure}[t]
  \centering
  \includegraphics[width=\columnwidth]
  {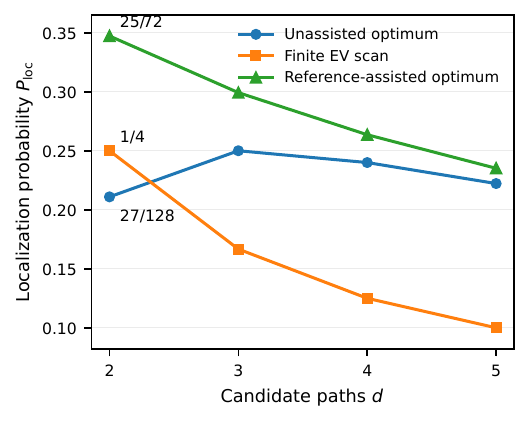}
  \caption{\label{fig:reference_rail_comparison}
  Lossless reference-rail advantage for one ideal absorber and two uses.
  The unassisted optimum acts only on the \(d\) candidate paths. The
  finite EV scan uses one promised empty rail in two successive balanced
  two-path tests. The reference-assisted optimum uses that same rail and
  is optimized over the full two-use causal class. The scan follows
  \(P_{\rm scan}^{(2)}=1/(2d)\). At \(d=2\), the three values are
  exactly \(27/128\), \(1/4\), and \(25/72\); the remaining points are
  numerical optima of the exact SDP, available in Table \ref{tab:app_reference_rail_lossless}. The reference-assisted optimum
  exceeds both alternatives in every solved dimension.}
\end{figure}

The exact hierarchy also reveals that the reference rail and the scan
must not be identified with one another. We therefore ask whether the
reference-assisted advantage is peculiar to two candidate paths.
Figure~\ref{fig:reference_rail_comparison} compares three lossless
\(K=2\) strategy classes for one ideal absorber and
\(d=2,\ldots,5\) candidate paths. The unassisted optimum uses only the
candidate paths. The finite EV scan uses one promised empty rail but
restricts the strategy to two successive balanced two-path tests. The
reference-assisted optimum uses that same rail and the same two
encounters, while optimizing over the full causal class.

The distinction becomes sharper as the path space grows. The rise of the unassisted curve from \(d=2\) to \(d=3\), followed by its decrease, reflects two competing effects: increasing \(d\) enlarges the dark subspace available to encode location, but it also increases the number of equiprobable absorber hypotheses. For the two-use data, the first effect dominates up to \(d=3\), while the second dominates thereafter. A two-use scan
tests only two candidate paths and therefore succeeds with
\(P_{\rm scan}^{(2)}(d,1)=1/(2d)\) for a uniformly distributed
absorber. It exceeds the unassisted optimum only at \(d=2\), where two
uses cover both candidates. For every solved dimension, however, the
fully optimized reference-assisted strategy exceeds both the scan and
the unassisted optimum. The exact \(d=2\) values are
\(27/128\), \(1/4\), and \(25/72\); the \(d=3,4,5\) points are
numerical optima of the corresponding exact semidefinite programs, collected in Table \ref{tab:app_reference_rail_lossless} (see Appendix \ref{app:reference_rail_benchmarks}).
Thus the promised rail remains useful beyond the minimal example, but
sequential scanning does not extract its full value.

\section{Loss, calibration, and practical realization}
\label{sec:implementation}
\subsection{Loss robustness}
An advantage of coherent multipath localization over
sequential interrogation appears under propagation loss, where a single
coherent pass pays the loss only once while a sequential scan compounds
it with depth. Let \(t=e^{-\alpha L}\) be the transmissivity of one
coherent pass through the \(d\)-path interferometer. Because the protocol
samples all \(d\) candidate locations in that single pass, the loss enters
as one factor \(t\) rather than accumulating across a chain of tests. Consider a multi-path recycling scheme, with
bright-port outcomes recycled back into the testing interferometer. The limiting probability that the first
non-bright termination is a correct dark-port localization event is
\begin{equation}
  \lambda_d^{\rm mp}(t)
  =
  \frac{d-1}{d}\,
  \frac{t(d-1)}{d^2-t(d-1)^2}.
  \label{eq:hd_loss_localization}
\end{equation}
The prefactor \((d-1)/d\) is the optimal dark-branch guessing
probability, and at \(t=1\) the expression reduces to the lossless value
\(\lambda_d=(d-1)^2/[d(2d-1)]\). Recycling raises the chance of reaching a
conclusive branch, while the limiting ratio is set by the one-pass
transmissivity and the multipath geometry.

\begin{figure*}[t]
  \centering
  \includegraphics[width=\textwidth]{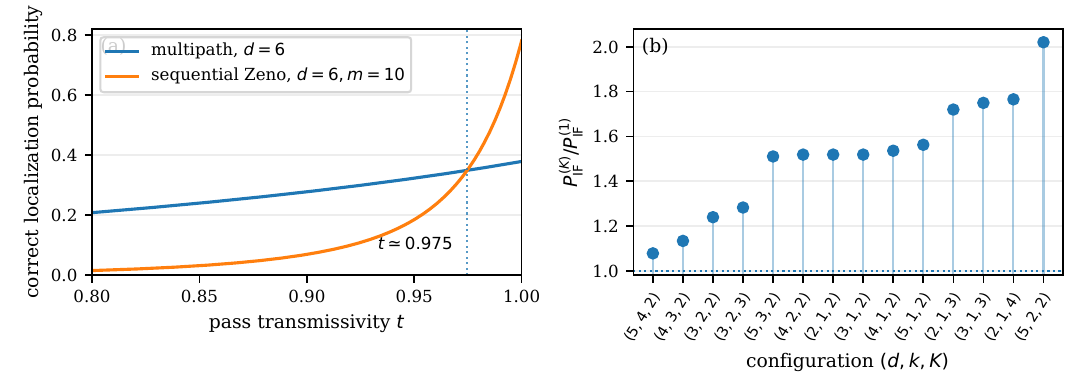}
  \caption{\label{fig:robustness_comparison}
  Calibrated robustness of interaction-free localization.
  (a) Loss-induced crossover between multipath localization and a sequential Zeno scan. The multipath curve is Eq.~\eqref{eq:hd_loss_localization} for \(d=6\); the Zeno curve is Eq.~\eqref{eq:zeno_loss_localization} with \(m=10\) recursions per candidate location. The dotted line marks the crossover.
  (b) Adaptive gain at \(t=0.9\) for all solved adaptive tester benchmarks in the candidate-path survival process. Each bar is labelled by \((d,k,K)\).
  }
\end{figure*}

A sequential Zeno scan obeys a different loss law.  If each
candidate location is tested by \(m\) weak Zeno recursions of
transmissivity \(t\), then, averaged over the unknown absorber position,
the correct-localization probability is
\begin{equation}
  \lambda_d^{\rm Zeno}(t,m)
  =
  \frac{t^m}{d}
  \cos^{2m}\!\left(\frac{\pi}{2m}\right)
  \frac{1-t^{md}}{1-t^m},
  \label{eq:zeno_loss_localization}
\end{equation}
with the \(t=1\) value understood by continuity. The Zeno factor
improves the lossless interrogation, but its transmissivity penalty is
paid at every weak recursion and across all earlier empty locations in the
scan, so loss compounds with depth and the two curves cross. For \(d=6\)
and \(m=10\) the crossover lies at \(t\simeq0.975\)
[Fig.~\ref{fig:robustness_comparison}(a)]: at \(t=0.95\) the multipath
protocol reaches a correct localization probability of \(0.32\), against
\(0.19\) for the EV-Zeno sequence, and at \(t=0.90\) the values are \(0.28\) and
\(0.07\), respectively. Below the crossover the single coherent pass wins precisely
because it avoids the depth penalty.

Throughout these comparisons a
success is a correct localization event, while absorption and
propagation loss count as failures. Eqs.~\eqref{eq:hd_loss_localization} and~\eqref{eq:zeno_loss_localization} are derived in Appendix \ref{app:loss_comparison}.

\subsection{Calibration and Device Imperfections}

For an experiment, the ideal projector absorber is replaced by a calibrated no-absorption process. The ideal maps
\(\mathcal S_j(\rho)=\oper{N_j}\rho\oper{N_j}\) become measured operations
\(\mathcal S_j^{\rm exp}\) with Choi operators
\(\sigma_j^{\rm exp}=J(\mathcal S_j^{\rm exp})\) describing the real behaviour of the interferometer when some of its paths are blocked. For independent uses the
\(K\)-use process is \(\Sigma_j^{\rm exp}=(\sigma_j^{\rm exp})^{\otimes K}\),
and if the object carries memory the measured \(K\)-slot no-absorption comb is used instead. The one-shot discrimination problem and the adaptive tester
benchmark keep the same form under this replacement, such that only the process
operators change.

The defining IFM condition is that the empty interferometer must
not produce a dark-port location signal. Uniform propagation loss and
detector inefficiency lower the rates but do not by themselves create a
false interaction-free event. Writing \(\mathcal E_{\varnothing}\)
for the calibrated empty interferometer, \(\rho_{\rm in}\) for the input
state, and \(\Pi_D\) for the projector onto the dark output modes,
\begin{equation}
  \epsilon_{\rm dark}
  =
  \operatorname{Tr}\!\left[
  \Pi_D\,\mathcal E_{\varnothing}(\rho_{\rm in})
  \right],
  \label{eq:empty_dark_leakage}
\end{equation}
measured with no absorber present, sets the background against
which dark-port localization events are certified, with detector dark
counts and electronic backgrounds folded into the same calibration. The
same tolerance can be imposed directly at the process level,
\begin{equation}
  \sum_i
  \operatorname{Tr}\!\left(T_i\Sigma_{\varnothing}^{\rm exp}\right)
  \le \epsilon ,
  \label{eq:calibrated_false_positive}
\end{equation}
the calibrated counterpart of the ideal empty-device condition in
Eq.~\eqref{eq:if_tester_cone}, bounding the total probability that the
empty device yields a conclusive report.

The adaptive benchmark provides a second robustness test under propagation loss. Whereas
the comparison in the previous subsection pits named strategies against one another, the tester
optimization ranges over all causal strategies for the same calibrated
process. At \(t=0.9\), every solved adaptive instance in
Fig.~\ref{fig:robustness_comparison}(b) still exceeds its single-use
benchmark, with gains ranging from about \(1.08\) for \((d,k,K)=(5,4,2)\)
to \(2.02\) for \((5,2,2)\), and higher-round examples such as \(1.77\)
for \((2,1,4)\) and \(1.75\) for \((3,1,3)\). Temporal wiring therefore
remains useful when each pass is lossy. The complete set of values is
tabulated in Table \ref{tab:app_adaptive_lossy_values} (see Appendix \ref{app:adaptive_tables_unassisted}).

This process-level formulation also absorbs the usual
imperfections. Phase drift, imperfect splitting ratios, mode mismatch,
finite extinction, detector inefficiency, residual dark leakage, and
mode-dependent absorber response all enter through the calibrated maps
\(\mathcal S_j^{\rm exp}\) and the empty process
\(\mathcal S_{\varnothing}^{\rm exp}\). In this manner, while ideal formulas give symmetric benchmarks, a calibrated experiment inherits benchmarks fixed by its own measured processes.

\subsection{Practical Realization}
The protocol asks little of the hardware beyond what
high-dimensional interferometry already provides. Single-pass localization requires
four ingredients: an input multiport that prepares the balanced
\(d\)-path superposition, the absorber region, the inverse multiport that
defines the bright output, and a final multiport that reads out the dark
subspace. Each ingredient is standard on current high-dimensional
photonic platforms. Candidate architectures, including path encodings in multicore fibers and integrated photonic
circuits, time-bin loop interferometers, and orbital
angular momentum modes, are capable of realizing the preparation, recombination, and readout
stages (see Ref.~\cite{MalikRoadmap} for a recent review).
Programmable universal processors have already demonstrated qudit
interaction-free interrogation at dimensions up to \(d=5\), so the
single-pass localization circuit is within reach of existing
devices~\cite{Giordani2023,Franco2026}.
The optimal readout is the
finite POVM that discriminates the output states with minimum-error, which can also be implemented by multipath interference with auxiliary modes \cite{Martinez2023Certification,martinez2025,yasir2025compactifyinglinearopticalunitaries}.  The collective readout
therefore remains realizable on the same hardware, with overhead that
grows only with the configuration to be resolved.

Adaptive strategies require the sequential/recycling architecture and might employ feedforward schemes previously demonstrated, which include low-latency switching and real-time control~\cite{LuizZanin:21,Duggan2026Feedforward}. The temporal wiring that lifts the
localization probability beyond the one-shot ceiling
[Fig.~\ref{fig:robustness_comparison}(b)] is thus an engineering of
existing components rather than a new device.

Although we have described an optical implementation for
concreteness, nothing in the construction is specific to photons. The
protocol needs only a coherent \(d\)-path interferometer, a localized
absorber on the candidate paths, and a path-resolving readout. These are
available in matter-wave and electron interferometry, in ultracold-atom
Zeno interrogation~\cite{Peise2015InteractionFree}, and at hard-x-ray
wavelengths where loss-resilient interaction-free schemes have been
analyzed~\cite{Cohen25}. In each case the dark branch plays the same
role: its location is carried entirely in the surviving quantum state,
and the absorber need not produce any classical signal during the run.

\section{Summary}
\label{sec:discussion}
Interaction-free measurement is usually read as a statement about presence. A dark-port click certifies that the empty interference pattern has been disturbed, while the detected photon has not been absorbed. The results above show that this is not the full content of the no-interaction branch. In a genuinely multipath interferometer, the dark output is a subspace. It can carry information about which path contained the absorber, and that information has an exact one-shot limit.

The exact optimum in Eq.~\eqref{eq:one_shot_optimal_value} is the main result. It is not tied to the balanced multiport as an assumed architecture. The optimization allows arbitrary probe states, arbitrary ancillas, and arbitrary measurements, subject only to the interaction-free constraint that no location is reported in the empty interferometer. The proof shows that ancillas do not enlarge the useful geometry. The only information available for localization is the absorber-induced component orthogonal to the empty-device state. The balanced multiport attains the resulting bound by producing the regular-simplex dark-port ensemble and reading it out with the corresponding square-root measurement.

This also clarifies the role of dimension. In two paths, the dark event remains binary. It can certify disturbance, but it does not provide a nontrivial one-shot location record. With three or more paths, the dark subspace has enough dimension to support distinct absorber-labelled states. The transition from two paths to many paths is therefore not only an efficiency change. It changes what the no-interaction branch can encode.

The same distinction appears in the comparison with sequential and Zeno-type schemes. Zeno interrogation improves a two-path test by making the interaction weak and repeated, usually with a known reference arm. Multipath localization addresses a different question: whether one coherent no-absorption branch can carry spatial information about an unknown absorber location. Sequential scans build localization from a chain of yes-or-no tests. The multipath protocol builds it directly into the dark-state geometry. In the presence of loss, this difference becomes operational, because a coherent multipath interrogation samples all candidate locations in one pass, whereas a sequential scan pays a depth cost before the relevant location is reached.

For several absorbers the no-interaction branch becomes a collective record: rather than reporting the first occupied path it encounters, the photon leaves the interferometer in a single state labelled by the occupied subset, with overlaps set entirely by the subset intersections. This is the natural many-path extension of the single-absorber simplex mechanism, and the resulting geometry lets the branch resolve subsets that a first-click scan cannot without adding extra ancillas. In symmetric cases such as \(d=4\), \(k=2\) it distinguishes any two path-sharing subsets perfectly, with only complementary subsets remaining degenerate. The point is not that every configuration is perfectly resolved, but that the branch stores relational information a sequential scan can only record with exponentially increasing resources.

The adaptive formulation separates the absorber hypotheses from the resources available to the optical strategy. The absorber configuration defines a process, and the optical apparatus defines a causal tester. Sequential scans, recycling circuits, Zeno-type protocols and adaptive multipath strategies are different testers for a fixed process model. The finite-round benchmark therefore asks for the best interaction-free localization strategy compatible with causal order, memory, feedforward and final measurement, with an optimal value depending on available resources. Solved instances of unassisted strategies (without a promised empty rail) show that temporal wiring can increase the localization probability beyond the exact single-use ceiling. Adding a free rail allows for reference-assisted strategies, which increase further the success probability. The exact two-use solution makes this distinction explicit. A second unassisted use raises the two-path optimum from $1/8$ to $27/128$; supplying the extra rail required by a sequential EV scan raises the value to $1/4$; and optimization over all two-use strategies with that same rail gives $25/72$. Sequential scanning is therefore not the optimal use of the reference-assisted geometry. For larger configurations, the numerical results show that the finite-round benchmark with access to an extra empty rail continues to raise the localization probability beyond the one achieved by unassisted strategies.

The robustness formulation is the experimental counterpart of this statement. A nonideal absorber should not be compressed into a single opacity parameter. On the non-absorbing branch it may attenuate, phase-shift, scatter, distort modes, or couple to inaccessible degrees of freedom. The object to calibrate is therefore the non-absorption subchannel itself. Empty-device dark leakage fixes the false-positive floor, the calibrated process fixes the information written into the surviving field, and pass transmissivity fixes the comparison with sequential interrogation. The ideal formulas are symmetric benchmarks; an experiment has its own benchmark determined by the measured processes. A natural next direction is to optimize strategies directly against such calibration data, for example through a tilted success-versus-background objective whose parameters encode the tolerated false-report cost, loss, and absorber contrast.

Together, these results identify a different use of IFM. The dark port is not only a herald that an object was present. In a multipath interferometer it is a quantum register, and the structure of that register determines how much spatial information can be extracted without absorption. The one-shot theorem gives the limit, a scheme built on balanced multiport beam splitters attains it, and the comb-tester formulation shows how the same task extends to finite-round adaptive strategies. Interaction-free measurement can therefore serve as a tool for object localization, not only for yes/no detection.

\begin{acknowledgments}
This research was funded by Fondo Nacional de Desarrollo
Cient\'{\i}fico y Tecnol\'ogico (FONDECYT) Grant Nos.\ 1240746, 1260111,
ANID -- Millennium Science Initiative Program -- ICN17$\_$012,
and ANID Anillo Project ATE250003. Additional support was provided by KLAR Grant No. BNI/PST/2023/1/00013/U/00001 funded by NAWA, and Narodowe Centrum Nauki under SONATINA project No. 2025/56/C/ST2/00058.
\end{acknowledgments}
\clearpage

\appendix

\section{Proof of the one-shot optimum}
\label{app:one_shot_proof}

This appendix gives the complete proof of the one-shot bound used in Theorem~\ref{thm:one_shot_optimality}. The proof separates three facts. First, the empty-interferometer constraint removes the no-absorber component from every conclusive outcome. Second, any ancilla-assisted probe induces the same informative Gram matrix as a signal-only probe with the same path weights. Third, a simple dual certificate bounds the resulting discrimination problem, and the uniform weights saturate the bound.

\subsection{From interaction-free localization to state discrimination}

Let the probe and ancilla be prepared in a pure state \(\ket{\Psi}_{SA}\). This entails no loss of generality. The feasible set is convex in the input state, the objective is linear, and any mixed state can be purified by enlarging the ancilla.

The empty-device condition in the main text is
\begin{equation}
  \bra{\Psi}M_a\ket{\Psi}=0
  \qquad \forall a .
  \label{eq:app_empty_expectation_zero}
\end{equation}
Since \(M_a\succeq0\), Eq.~\eqref{eq:app_empty_expectation_zero} is equivalent to
\begin{equation}
  M_a\ket{\Psi}=0
  \qquad \forall a .
  \label{eq:app_empty_vector_zero}
\end{equation}
Indeed, \(\bra{\Psi}M_a\ket{\Psi}=\|M_a^{1/2}\ket{\Psi}\|^2\), so the expectation vanishes exactly when \(M_a^{1/2}\ket{\Psi}=0\), and hence when \(M_a\ket{\Psi}=0\).

Define the projector onto the subspace orthogonal to the empty-device state,
\begin{equation}
  Q=I-\ket{\Psi}\bra{\Psi} .
  \label{eq:app_empty_orthogonal_projector}
\end{equation}
For an absorber in path \(a\), the no-absorption operator is \(N_a=I-\ket a\bra a\). The only part of \((N_a\otimes I_A)\ket{\Psi}\) that can contribute to the conclusive effect \(M_a\) is its component orthogonal to \(\ket{\Psi}\). We therefore define
\begin{equation}
  \ket{w_a}
  =
  Q(N_a\otimes I_A)\ket{\Psi} .
  \label{eq:app_informative_vector}
\end{equation}
Using Eq.~\eqref{eq:app_empty_vector_zero}, the success probability becomes
\begin{equation}
  P_{\rm loc}
  =
  \frac1d\sum_{a=0}^{d-1}
  \bra{w_a}M_a\ket{w_a} .
  \label{eq:app_success_w_vectors}
\end{equation}
Thus the interaction-free localization task has become minimum-error discrimination of the subnormalized pure states
\begin{equation}
  \tau_a=\frac1d\ket{w_a}\bra{w_a} .
  \label{eq:app_tau_states}
\end{equation}
The empty branch has not disappeared; it is precisely what forces the projection in Eq.~\eqref{eq:app_informative_vector}.

\subsection{Eliminating the ancilla}

Write the pure probe-ancilla state as
\begin{equation}
  \ket{\Psi}
  =
  \sum_{i=0}^{d-1}\ket{i}\ket{\eta_i},
  \qquad
  p_i=\langle\eta_i|\eta_i\rangle,
  \qquad
  \sum_i p_i=1 .
  \label{eq:app_purified_probe_weights}
\end{equation}
The number \(p_i\) is the weight placed on path \(i\). Since
\begin{equation}
  (N_a\otimes I_A)\ket{\Psi}
  =
  \ket{\Psi}-\ket a\ket{\eta_a},
  \label{eq:app_absorber_removed_component}
\end{equation}
projection with \(Q\) gives
\begin{equation}
  \ket{w_a}
  =
  p_a\ket{\Psi}-\ket a\ket{\eta_a} .
  \label{eq:app_w_explicit}
\end{equation}
The Gram matrix of the informative vectors is therefore
\begin{align}
  \langle w_a|w_a\rangle &= p_a(1-p_a),
  \label{eq:app_w_norm}\\
  \langle w_a|w_b\rangle &= -p_ap_b,
  \qquad a\neq b .
  \label{eq:app_w_overlap}
\end{align}
The derivation uses only the path weights \(p_i\), not the overlaps among the ancilla states \(\ket{\eta_i}\).

Now define the signal-only state with the same path weights,
\begin{equation}
  \ket{\psi_p}
  =
  \sum_{i=0}^{d-1}\sqrt{p_i}\ket{i},
  \label{eq:app_signal_only_probe}
\end{equation}
and the corresponding projector \(Q_p=I-\ket{\psi_p}\bra{\psi_p}\). Its informative vectors are
\begin{equation}
  \ket{u_a}=Q_pN_a\ket{\psi_p}
  =
  p_a\ket{\psi_p}-\sqrt{p_a}\ket a .
  \label{eq:app_u_explicit}
\end{equation}
They satisfy
\begin{align}
  \langle u_a|u_a\rangle &= p_a(1-p_a),
  \label{eq:app_u_norm}\\
  \langle u_a|u_b\rangle &= -p_ap_b,
  \qquad a\neq b .
  \label{eq:app_u_overlap}
\end{align}
Equations~\eqref{eq:app_w_norm}-\eqref{eq:app_w_overlap} and \eqref{eq:app_u_norm}-\eqref{eq:app_u_overlap} are identical. Hence the two informative ensembles are isometric. Minimum-error discrimination of pure-state ensembles depends only on the Gram matrix, so every ancilla-assisted strategy has a signal-only strategy with the same value. Ancillas therefore do not improve the one-shot optimum.

\subsection{Dual certificate for fixed path weights}

Fix the probability vector \(p=(p_0,\ldots,p_{d-1})\). Let \(A\) be the linear map whose \(a\)-th column is \(\ket{u_a}\), and define
\begin{equation}
  W=\frac1dAA^\dagger
  =
  \frac1d\sum_{a=0}^{d-1}\ket{u_a}\bra{u_a} .
  \label{eq:app_W_operator}
\end{equation}
The primal minimum-error discrimination problem is
\begin{equation}
  \max_{\{M_a\}}
  \sum_a \operatorname{Tr}(M_a\tau_a),
  \qquad
  \sum_a M_a\le I,
  \qquad
  M_a\succeq0 .
  \label{eq:app_discrimination_primal}
\end{equation}
The dual problem is
\begin{equation}
  \min_Y \operatorname{Tr}Y,
  \qquad
  Y\succeq \tau_a\quad \forall a .
  \label{eq:app_discrimination_dual}
\end{equation}
Thus any positive operator \(Y\) dominating every \(\tau_a\) gives an upper bound.

The columns of \(A\) sum to zero, because
\begin{equation}
  \sum_a\ket{u_a}
  =
  \left(\sum_a p_a\right)\ket{\psi_p}
  -
  \sum_a\sqrt{p_a}\ket a
  =0 .
  \label{eq:app_columns_sum_zero}
\end{equation}
Therefore the all-ones vector belongs to the kernel of \(A\). Let
\begin{equation}
  P_{\rm row}=A^\dagger(AA^\dagger)^+A
  \label{eq:app_row_projector}
\end{equation}
be the orthogonal projector onto the row space of \(A\), where \((\cdot)^+\) denotes the Moore--Penrose inverse. Since the row space is contained in the subspace orthogonal to the all-ones vector,
\begin{equation}
  P_{\rm row}\preceq I-\frac1dJ,
  \label{eq:app_row_projector_bound}
\end{equation}
where \(J\) is the all-ones matrix. Taking the \(a\)-th diagonal element gives
\begin{equation}
  \langle e_a|P_{\rm row}|e_a\rangle\le \frac{d-1}{d} .
  \label{eq:app_row_diagonal_bound}
\end{equation}

We use the following elementary domination fact: if \(M\succeq0\) and \(v\in{\rm Ran}(M)\), then
\begin{equation}
  \ket v\bra v
  \preceq
  \langle v|M^+|v\rangle M .
  \label{eq:app_rank_one_domination}
\end{equation}
To see this, restrict to the support of \(M\), write \(v=M^{1/2}y\), and apply Cauchy's inequality to \(|\langle x|v\rangle|^2=|\langle M^{1/2}x|y\rangle|^2\).

Apply Eq.~\eqref{eq:app_rank_one_domination} with \(M=AA^\dagger\) and \(v=\ket{u_a}=A\ket{e_a}\). Then
\begin{equation}
  \ket{u_a}\bra{u_a}
  \preceq
  \langle e_a|P_{\rm row}|e_a\rangle AA^\dagger
  \preceq
  \frac{d-1}{d}AA^\dagger .
  \label{eq:app_u_domination}
\end{equation}
Dividing by \(d\) and using \(W=AA^\dagger/d\), we obtain
\begin{equation}
  \tau_a
  =
  \frac1d\ket{u_a}\bra{u_a}
  \preceq
  \frac{d-1}{d}W
  \qquad \forall a .
  \label{eq:app_dual_feasible}
\end{equation}
Thus
\begin{equation}
  Y=\frac{d-1}{d}W
  \label{eq:app_dual_certificate}
\end{equation}
is dual feasible. Its trace gives
\begin{align}
  P_{\rm loc}(p)
  &\le
  \operatorname{Tr}Y
  =
  \frac{d-1}{d}\operatorname{Tr}W
  \nonumber\\
  &=
  \frac{d-1}{d^2}
  \left(1-\sum_{a=0}^{d-1}p_a^2\right).
  \label{eq:app_path_weight_bound}
\end{align}
This is the path-weight bound used in the main text.

\subsection{Optimization over path weights and saturation}

The term \(\sum_a p_a^2\) is minimized by the uniform distribution. Therefore
\begin{equation}
  1-\sum_a p_a^2
  \le
  1-\frac1d
  =
  \frac{d-1}{d},
  \label{eq:app_uniform_weight_bound}
\end{equation}
with equality if and only if \(p_a=1/d\) for all \(a\). Combining Eqs.~\eqref{eq:app_path_weight_bound} and \eqref{eq:app_uniform_weight_bound} gives
\begin{equation}
  P_{\rm loc}
  \le
  \frac{(d-1)^2}{d^3} .
  \label{eq:app_one_shot_global_bound}
\end{equation}

It remains to check that the bound is attainable. For the uniform probe,
\begin{equation}
  \ket{u_a}
  =
  \frac1d\ket{\psi_{\rm bal}}
  -
  \frac1{\sqrt d}\ket a .
  \label{eq:app_uniform_u_vectors}
\end{equation}
The norms and overlaps are
\begin{equation}
  \|u_a\|^2=\frac{d-1}{d^2},
  \qquad
  \langle u_a|u_b\rangle=-\frac1{d^2}
  \quad (a\neq b).
  \label{eq:app_uniform_u_gram}
\end{equation}
After normalization, the \(d\) states form a regular simplex in a \((d-1)\)-dimensional subspace. The square-root measurement is optimal for this symmetric ensemble and has conditional success probability \((d-1)/d\). Since the dark informative branch occurs with probability \((d-1)/d^2\), the total success probability is
\begin{equation}
  \frac{d-1}{d^2}\cdot\frac{d-1}{d}
  =
  \frac{(d-1)^2}{d^3} .
  \label{eq:app_saturation_value}
\end{equation}
This saturates Eq.~\eqref{eq:app_one_shot_global_bound} and completes the proof of the one-shot optimum.

\section{Multiple-absorber state geometry}
\label{app:k_absorbers}

This appendix records the geometry behind the subset-labelled dark states used in the many-absorber section. The derivation is included so that the many-absorber statement is not a separate assumption but the same multiport mechanism applied to a higher-rank absorber projector.

Let \(\mathcal A\subset\{0,\ldots,d-1\}\) be the occupied set, with \(|\mathcal A|=k\). The absorber projector and no-absorption operator are
\begin{equation}
  A_{\mathcal A}=\sum_{a\in\mathcal A}\ket a\bra a,
  \qquad
  N_{\mathcal A}=I-A_{\mathcal A} .
  \label{eq:app_subset_projectors}
\end{equation}
The balanced input is \(d^{-1/2}\sum_i\ket i\). Therefore the absorption probability is the total weight on the occupied paths,
\begin{equation}
  P_*^{(k)}=\frac{k}{d} .
  \label{eq:app_k_absorption}
\end{equation}
After recombination by the inverse multiport, the surviving state is
\begin{equation}
  \ket{\psi_e^{\mathcal A}}
  =
  \ket0
  -
  \frac1d
  \sum_{a\in\mathcal A}\sum_{n=0}^{d-1}u_{an}^*\ket n .
  \label{eq:app_k_exit_state}
\end{equation}
The bright amplitude is the coefficient of \(\ket0\). Since \(u_{a0}=1\), it equals \(1-k/d=(d-k)/d\), hence
\begin{equation}
  P_b^{(k)}=\left(\frac{d-k}{d}\right)^2 .
  \label{eq:app_k_bright}
\end{equation}
The total survival probability is \((d-k)/d\). Subtracting the bright probability gives the dark probability
\begin{equation}
  P_s^{(k)}
  =
  \frac{d-k}{d}
  -
  \left(\frac{d-k}{d}\right)^2
  =
  \frac{k(d-k)}{d^2} .
  \label{eq:app_k_dark}
\end{equation}
The corresponding interaction-free detection efficiency is
\begin{equation}
  \eta_{d,k}
  =
  \frac{P_s^{(k)}}{P_s^{(k)}+P_*^{(k)}}
  =
  \frac{d-k}{2d-k} .
  \label{eq:app_k_efficiency}
\end{equation}

Conditioned on a dark event, the normalized state is
\begin{equation}
  \ket{\phi_{s\mathcal A}}
  =
  -\frac1{\sqrt{k(d-k)}}
  \sum_{a\in\mathcal A}\sum_{n=1}^{d-1}u_{an}^*\ket n .
  \label{eq:app_k_dark_state}
\end{equation}
To compute the overlap of two such states, use the dephased Hadamard identity
\begin{equation}
  \sum_{n=1}^{d-1}u_{an}u_{a'n}^*
  =
  \begin{cases}
  d-1, & a=a',\\
  -1, & a\neq a'.
  \end{cases}
  \label{eq:app_dephased_identity}
\end{equation}
If \(x=|\mathcal A\cap\mathcal A'|\), then there are \(x\) equal-index pairs and \(k^2-x\) unequal-index pairs. Hence
\begin{align}
  \langle\phi_{s\mathcal A'}|\phi_{s\mathcal A}\rangle
  &=
  \frac{x(d-1)-(k^2-x)}{k(d-k)}
  \nonumber\\
  &=
  \frac{dx-k^2}{k(d-k)} .
  \label{eq:app_k_overlap_signed}
\end{align}
The magnitude is the overlap formula used in the main text.

For uniform priors and unit relevance weights, the conditional readout on the dark branch is the finite minimum-error discrimination problem
\begin{align}
  G_k^{\rm opt}
  =
  \max_{\{\Pi_{\mathcal A}\}}
  &\frac1{\binom dk}
  \sum_{|\mathcal A|=k}
  \operatorname{Tr}(\Pi_{\mathcal A}\rho_{\mathcal A}),
  \label{eq:app_k_discrimination}\\
  \mathrm{subject\ to}\quad
  &\sum_{|\mathcal A|=k}\Pi_{\mathcal A}=I,
  \qquad
  \Pi_{\mathcal A}\succeq0 .
  \nonumber
\end{align}
Here \(\rho_{\mathcal A}=\ket{\phi_{s\mathcal A}}\bra{\phi_{s\mathcal A}}\). More general priors or relevance weights only change the linear objective, not the state geometry.

The uniform subset ensemble has an analytic optimum. Write the single-absorber dark states as \(\{\ket{\phi_a}\}_{a=0}^{d-1}\), with
\begin{equation}
  \sum_{a=0}^{d-1}\ket{\phi_a}=0,
  \qquad
  \sum_{a=0}^{d-1}\ket{\phi_a}\bra{\phi_a}
  =
  \frac{d}{d-1}I_D ,
  \label{eq:app_single_absorber_tight_frame}
\end{equation}
where \(I_D\) is the identity on the \((d-1)\)-dimensional dark subspace. The subset state can be written as
\begin{equation}
  \ket{\phi_{s\mathcal A}}
  =
  \sqrt{\frac{d-1}{k(d-k)}}
  \sum_{a\in\mathcal A}\ket{\phi_a}.
  \label{eq:app_subset_as_simplex_sum}
\end{equation}
Summing the projectors over all subsets gives
\begin{equation}
  \sum_{|\mathcal A|=k}\rho_{\mathcal A}
  =
  \frac{\binom dk}{d-1}I_D .
  \label{eq:app_subset_tight_frame}
\end{equation}
Indeed, each diagonal term \(\ket{\phi_a}\bra{\phi_a}\) appears in \(\binom{d-1}{k-1}\) subsets, each off-diagonal term \(\ket{\phi_a}\bra{\phi_b}\) with \(a\neq b\) appears in \(\binom{d-2}{k-2}\) subsets, and \(\sum_a\ket{\phi_a}=0\) converts the off-diagonal sum into minus the diagonal sum.

Equation~\eqref{eq:app_subset_tight_frame} makes the square-root measurement explicit:
\begin{equation}
  \Pi_{\mathcal A}
  =
  \frac{d-1}{\binom dk}\rho_{\mathcal A}.
  \label{eq:app_subset_srm}
\end{equation}
These operators are positive and sum to \(I_D\). Their success probability is
\begin{equation}
  G_k^{\rm opt}
  =
  \frac1{\binom dk}
  \sum_{|\mathcal A|=k}
  \operatorname{Tr}(\Pi_{\mathcal A}\rho_{\mathcal A})
  =
  \frac{d-1}{\binom dk}.
  \label{eq:app_subset_analytic_success}
\end{equation}
This value is optimal because \(Y=I_D/\binom dk\) is dual feasible: for every subset,
\(Y\succeq \rho_{\mathcal A}/\binom dk\), and \(\operatorname{Tr}Y=(d-1)/\binom dk\). Thus the balanced-architecture total exact-subset localization probability is
\begin{equation}
  P_{\rm loc}^{(d,k)}
  =
  P_s^{(k)}G_k^{\rm opt}
  =
  \frac{k(d-k)}{d^2}\,
  \frac{d-1}{\binom dk}.
  \label{eq:app_k_total_analytic}
\end{equation}

The simplest nontrivial example is \(d=4,k=2\). If two subsets share one path, then \(x=1\) and Eq.~\eqref{eq:app_k_overlap_signed} gives zero overlap. If they are complementary, then \(x=0\) and the magnitude is one. Thus the six subsets collapse into three orthogonal rays, paired by complementation:
\begin{equation}
  \{01,23\},\qquad \{02,13\},\qquad \{03,12\} .
  \label{eq:app_d4k2_pairs}
\end{equation}
A single dark-branch photon can distinguish which complementary pair occurred, while no measurement can distinguish the two complementary subsets within a pair. This illustrates the point made in the main text: the multipath branch stores collective subset information, including relational information.

\begin{table*}[t]
\centering
\begin{tabular}{c c c c c c c}
\hline\hline
\((d,k)\) & configurations & \(P_*\) & \(P_s\) & \(P_b\) & conditional readout & total one-shot value \\
\hline
  $(2,1)$ & 2 & 0.500000 & 0.250000 & 0.250000 & 0.500000 & 0.125000 \\
  $(3,1)$ & 3 & 0.333333 & 0.222222 & 0.444444 & 0.666667 & 0.148148 \\
  $(3,2)$ & 3 & 0.666667 & 0.222222 & 0.111111 & 0.666667 & 0.148148 \\
  $(4,1)$ & 4 & 0.250000 & 0.187500 & 0.562500 & 0.750000 & 0.140625 \\
  $(4,2)$ & 6 & 0.500000 & 0.250000 & 0.250000 & 0.500000 & 0.125000 \\
  $(4,3)$ & 4 & 0.750000 & 0.187500 & 0.062500 & 0.750000 & 0.140625 \\
  $(5,1)$ & 5 & 0.200000 & 0.160000 & 0.640000 & 0.800000 & 0.128000 \\
  $(5,2)$ & 10 & 0.400000 & 0.240000 & 0.360000 & 0.400000 & 0.096000 \\
  $(5,3)$ & 10 & 0.600000 & 0.240000 & 0.160000 & 0.400000 & 0.096000 \\
  $(5,4)$ & 5 & 0.800000 & 0.160000 & 0.040000 & 0.800000 & 0.128000 \\
\hline\hline
\end{tabular}
\caption{Small balanced-architecture values for representative \((d,k)\). The probabilities \(P_*\), \(P_s\), and \(P_b\) follow from Eqs.~\eqref{eq:app_k_absorption}-\eqref{eq:app_k_dark}. The conditional readout column is the optimal minimum-error discrimination value of the subset-labelled dark states for the listed symmetric cases, computed from Eq.~\eqref{eq:app_k_discrimination}. The final column is \(P_s\) multiplied by the conditional readout value.}
\label{tab:app_exact_small_values}
\end{table*}

\section{Calibrated processes and adaptive testers}
\label{app:calibrated_testers}

This appendix gives the process-level formulation used in the adaptive and robustness sections. The point is to separate the object from the strategy. The object is the calibrated process. The strategy is a tester. Their interface is a trace pairing.

\subsection{One use of a general absorber}

Let \(A\) be the Hilbert space entering the object region and \(B\) the Hilbert space returning from it on the non-absorbing branch. These spaces may include path, polarization, frequency, temporal mode, spatial mode, and accessible auxiliary output modes. For absorber configuration \(j\), the non-absorption subchannel is a completely positive trace-nonincreasing map
\begin{align}
  \mathcal S_j^{\rm exp}&:\mathcal L(A)\to\mathcal L(B),
  \nonumber\\
  \mathcal S_j^{\rm exp}(\rho)
  &=
  \sum_\mu K_{j\mu}\rho K_{j\mu}^\dagger,
  \qquad
  \sum_\mu K_{j\mu}^\dagger K_{j\mu}\le I_A .
  \label{eq:app_no_click_map}
\end{align}
The operator
\begin{equation}
  F_j=\sum_\mu K_{j\mu}^\dagger K_{j\mu}
  \label{eq:app_no_click_effect}
\end{equation}
is the non-absorption effect. For input \(\rho\), the no-absorption exit probability is
\begin{equation}
  p_{\rm e}(j|\rho)
  =
  \operatorname{Tr}[\mathcal S_j^{\rm exp}(\rho)]
  =
  \operatorname{Tr}(F_j\rho).
  \label{eq:app_no_click_probability}
\end{equation}
The ideal absorber used in the exact one-shot theorem is the special case
\begin{equation}
  \mathcal S_j(\rho)=\bar P_j\rho\bar P_j,
  \qquad
  \bar P_j=I-P_j .
  \label{eq:app_ideal_projector_process}
\end{equation}
Equation~\eqref{eq:app_no_click_map} is therefore not a scalar-opacity correction. It allows finite opacity, phase shifts, scattering into auxiliary modes, polarization or frequency distortion, mode mismatch, and detector internal degrees of freedom.

\subsection{Choi representation}
\label{app:Choi_rep}

Choose an orthonormal basis \(\{\ket\alpha\}\) for \(A\), and define the unnormalized maximally entangled vector
\begin{equation}
  \ket\Phi_{AA'}=\sum_\alpha\ket\alpha_A\ket\alpha_{A'} .
  \label{eq:app_phi}
\end{equation}
For a map \(\mathcal M:A\to B\), define
\begin{equation}
  J(\mathcal M)
  =
  (\mathcal M\otimes {\rm id}_{A'})(\ket\Phi\bra\Phi).
  \label{eq:app_choi}
\end{equation}
After identifying \(A'\simeq A\), the Choi operator acts on \(B\otimes A\). Expanding the definition gives
\begin{equation}
  J(\mathcal M)
  =
  \sum_{\alpha,\beta}\mathcal M(\ket\alpha\bra\beta)\otimes\ket\alpha\bra\beta .
  \label{eq:app_choi_expansion}
\end{equation}
The map is recovered by contracting the input leg with a transpose:
\begin{equation}
  \mathcal M(X)
  =
  \operatorname{Tr}_A
  \left[
  J(\mathcal M)(I_B\otimes X^T)
  \right].
  \label{eq:app_choi_recovery}
\end{equation}
The transpose is taken in the same basis used in Eq.~\eqref{eq:app_phi}. To verify Eq.~\eqref{eq:app_choi_recovery}, write \(X=\sum_{\mu,\nu}X_{\mu\nu}\ket\mu\bra\nu\). The trace over \(A\) pairs \(\ket\alpha\bra\beta\) with \(\ket\nu\bra\mu\), leaving exactly \(\sum_{\mu,\nu}X_{\mu\nu}\mathcal M(\ket\mu\bra\nu)=\mathcal M(X)\).

For the calibrated processes write
\begin{equation}
  \sigma_j^{\rm exp}=J(\mathcal S_j^{\rm exp}) .
  \label{eq:app_sigma_exp}
\end{equation}
Complete positivity is equivalent to
\begin{equation}
  \sigma_j^{\rm exp}\succeq0,
  \label{eq:app_sigma_positive}
\end{equation}
and trace nonincrease is equivalent to
\begin{equation}
  \operatorname{Tr}_B\sigma_j^{\rm exp}=F_j^T\le I_A .
  \label{eq:app_trace_nonincrease}
\end{equation}
For the ideal projector model, the Choi operator is rank one,
\begin{equation}
  \sigma_j=\ket{\bar P_j}\!\rangle\langle\!\bra{\bar P_j},
  \label{eq:app_rank_one_projector_choi}
\end{equation}
because the no-absorption map has one Kraus operator. A general calibrated absorber need not have this rank-one structure.

\subsection{The \texorpdfstring{\(K\)}{K}-use object process}

For \(K\) uses, introduce ordered slots \(A_r\to B_r\), \(r=1,\ldots,K\). If the same calibrated operation is inserted independently at each use, then
\begin{equation}
  \Sigma_j^{\rm exp}
  =
  (\sigma_j^{\rm exp})^{\otimes K} .
  \label{eq:app_product_process}
\end{equation}
This tensor product is a statement about the object. It does not impose a memoryless strategy. If the object or detector has memory across uses, \(\Sigma_j^{\rm exp}\) is instead the measured \(K\)-slot no-click comb on \(B_KA_K\cdots B_1A_1\).

\subsection{Adaptive strategies and tester elements}
\label{app:adaptive_strategies}

A general adaptive strategy begins with a state on \(R_0A_1\), applies channels
\begin{equation}
  \Lambda_r:B_rR_{r-1}\to R_rA_{r+1},
  \qquad r=1,\ldots,K-1,
  \label{eq:app_internal_channels}
\end{equation}
and ends with a measurement on \(B_KR_{K-1}\). Intermediate measurements and feedforward are included because a measurement followed by a classically controlled operation is a channel that writes the outcome into a memory register.

For each conclusive final report \(i\), the fixed parts of the adaptive circuit contract to a positive operator
\begin{equation}
  T_i\succeq0,
  \qquad
  T_i\in\mathcal L(B_KA_K\cdots B_1A_1),
  \label{eq:app_tester_element}
\end{equation}
called a tester element. If the true process is \(\Sigma_j^{\rm exp}\), then
\begin{equation}
  p(i,{\rm e}|j)
  =
  \operatorname{Tr}(T_i\Sigma_j^{\rm exp}) .
  \label{eq:app_trace_pairing}
\end{equation}
This is the basic probability identity: the tester describes the strategy branch, and the process describes the interrogated object.

Let
\begin{equation}
  \Omega=\sum_{i=1}^N T_i+T_\varnothing
  \label{eq:app_deterministic_tester}
\end{equation}
be the deterministic tester obtained by ignoring the final outcome. A valid \(K\)-round tester is characterized by positive operators \(Q_1,\ldots,Q_K\) satisfying
\begin{align}
  \Omega&=I_{B_K}\otimes Q_K,
  \label{eq:app_tester_norm_1}\\
  \operatorname{Tr}_{A_r}Q_r
  &=
  I_{B_{r-1}}\otimes Q_{r-1},
  \qquad r=2,\ldots,K,
  \label{eq:app_tester_norm_2}\\
  \operatorname{Tr}_{A_1}Q_1&=1 .
  \label{eq:app_tester_norm_3}
\end{align}
Equation~\eqref{eq:app_tester_norm_1} says that after the last output is received, ignoring the final measurement leaves identity on the last output. Equation~\eqref{eq:app_tester_norm_2} is the trace-preserving condition for each internal update. Equation~\eqref{eq:app_tester_norm_3} normalizes the initial preparation. Together they express causal order: earlier outputs may affect later inputs, but later outputs cannot affect earlier inputs.

\subsection{Ideal and calibrated interaction-free constraints}

Let \(\Sigma_\varnothing\) be the empty-device process. In the ideal IFM problem, no conclusive location report is allowed in the empty device:
\begin{equation}
  \sum_{i=1}^N\operatorname{Tr}(T_i\Sigma_\varnothing)=0 .
  \label{eq:app_ideal_empty_constraint}
\end{equation}
All terms are nonnegative, so this is equivalent to \(\operatorname{Tr}(T_i\Sigma_\varnothing)=0\) for each \(i\). If \(\Sigma_\varnothing=\ket{v_\varnothing}\bra{v_\varnothing}\), then for \(T_i\succeq0\),
\begin{equation}
  \operatorname{Tr}(T_i\Sigma_\varnothing)=0
  \quad\Longleftrightarrow\quad
  T_i\ket{v_\varnothing}=0 .
  \label{eq:app_null_face}
\end{equation}
This is the null-face form used in the ideal numerical implementation.

For a calibrated device, exact zero background is replaced by a measured tolerance:
\begin{equation}
  \sum_{i=1}^N\operatorname{Tr}(T_i\Sigma_\varnothing^{\rm exp})\le \epsilon .
  \label{eq:app_calibrated_empty_constraint}
\end{equation}
This is the process-level version of empty-device dark leakage.

\subsection{Calibrated adaptive benchmark and exactness}

For a tolerated empty-device conclusive background \(\epsilon\), the calibrated adaptive value is
\begin{widetext}
\begin{align}
  P_\epsilon^{(K)}
  =
  \max_{\{T_j\},\Omega,\{Q_r\}}
  \quad&
  \frac1N\sum_{j=1}^N\operatorname{Tr}(T_j\Sigma_j^{\rm exp})
  \label{eq:app_calibrated_benchmark}\\
  \mathrm{subject\ to}\quad&
  T_j\succeq0\quad(j=1,\ldots,N),\qquad
  \Omega-\sum_{j=1}^N T_j\succeq0,
  \nonumber\\
  &
  \Omega=I_{B_K}\otimes Q_K,
  \nonumber\\
  &
  \operatorname{Tr}_{A_r}Q_r=I_{B_{r-1}}\otimes Q_{r-1},
  \qquad r=2,\ldots,K,
  \nonumber\\
  &
  \operatorname{Tr}_{A_1}Q_1=1,
  \qquad Q_r\succeq0,
  \nonumber\\
  &
  \sum_{j=1}^N\operatorname{Tr}(T_j\Sigma_\varnothing^{\rm exp})\le \epsilon .
  \nonumber
\end{align}
\end{widetext}
Every physical adaptive optical strategy gives feasible tester elements satisfying these constraints and has the objective value shown in Eq.~\eqref{eq:app_calibrated_benchmark}. Conversely, every feasible tester has a sequential realization. The equations for \(Q_r\) reconstruct the initial preparation and the deterministic internal updates. The decomposition of \(\Omega\) into positive pieces \(T_1,\ldots,T_N,T_\varnothing\) is implemented as a final measurement on the support of \(\Omega\). Hence Eq.~\eqref{eq:app_calibrated_benchmark} optimizes over exactly the adaptive strategies compatible with the calibrated processes.

The product form \(\Sigma_j=(\sigma_j)^{\otimes K}\) therefore does not remove adaptivity. It says only that the object process has no memory. The tester element \(T_i\) is a general operator on \(B_KA_K\cdots B_1A_1\), and its correlations represent memory, intermediate measurements, and feedforward.

\subsection{Reduction to the one-shot constraint}

At \(K=1\), an input \(\rho\) and final conclusive effect \(M_i\) define
\begin{equation}
  T_i=M_i\otimes\rho^T .
  \label{eq:app_one_shot_tester}
\end{equation}
For the ideal process \(\mathcal S_j(\rho)=\bar P_j\rho\bar P_j\),
\begin{equation}
  \operatorname{Tr}(T_i\sigma_j)
  =
  \operatorname{Tr}[M_i\mathcal S_j(\rho)] .
  \label{eq:app_one_shot_pairing}
\end{equation}
For the empty process \(\mathcal S_\varnothing(\rho)=\rho\),
\begin{equation}
  \operatorname{Tr}(T_i\sigma_\varnothing)
  =
  \operatorname{Tr}(M_i\rho).
  \label{eq:app_one_shot_empty}
\end{equation}
The ideal no-false-report condition is therefore exactly the one-shot empty-interferometer constraint.

\section{Adaptive benchmark data and verification}
\label{app:adaptive_tables}

This appendix lists the numerical values behind the adaptive figures. The numerical semidefinite programs reported below were solved with MOSEK. The columns report the optimized adaptive value of the success probability, the corresponding one-use value at the same transmissivity, and the gain over that one-use value.

\subsection{Unassisted benchmarks}
\label{app:adaptive_tables_unassisted}

\begin{table*}[t]
\centering
\begin{tabular}{c c c c c}
\hline\hline
\((d,k,K)\) & \(t\) & \(P_{\rm IF}^{(K)}\) & one-use value & gain \\
\hline
  $(2,1,1)$ & 1.0 & 0.125000 & 0.125000 & 1.000000 \\
  $(2,1,2)$ & 1.0 & 0.210938 & 0.125000 & 1.687500 \\
  $(2,1,3)$ & 1.0 & 0.265395 & 0.125000 & 2.123160 \\
  $(2,1,4)$ & 1.0 & 0.302714 & 0.125000 & 2.421714 \\
  $(3,1,1)$ & 1.0 & 0.148148 & 0.148148 & 1.000000 \\
  $(3,1,2)$ & 1.0 & 0.250000 & 0.148148 & 1.687500 \\
  $(3,1,3)$ & 1.0 & 0.319997 & 0.148148 & 2.159977 \\
  $(3,2,2)$ & 1.0 & 0.203997 & 0.148148 & 1.376981 \\
  $(4,1,2)$ & 1.0 & 0.240002 & 0.140625 & 1.706683 \\
  $(4,2,2)$ & 1.0 & 0.210939 & 0.125000 & 1.687511 \\
  $(4,3,2)$ & 1.0 & 0.177089 & 0.140625 & 1.259301 \\
  $(5,1,2)$ & 1.0 & 0.222214 & 0.128000 & 1.736044 \\
  $(5,2,2)$ & 1.0 & 0.215491 & 0.096000 & 2.244701 \\
  $(5,3,2)$ & 1.0 & 0.161123 & 0.096000 & 1.678362 \\
  $(5,4,2)$ & 1.0 & 0.153296 & 0.128000 & 1.197625 \\
\hline\hline
\end{tabular}
\caption{Ideal candidate-path adaptive benchmarks used in Fig.~\ref{fig:adaptive_advantage}.}
\label{tab:app_adaptive_ideal_values}
\end{table*}

\begin{table*}[t]
\centering
\begin{tabular}{c c c c c}
\hline\hline
\((d,k,K)\) & \(t\) & \(P_{\rm IF}^{(K)}\) & one-use value & gain \\
\hline
  $(5,4,2)$ & 0.9 & 0.124158 & 0.115200 & 1.077762 \\
  $(4,3,2)$ & 0.9 & 0.143442 & 0.126562 & 1.133371 \\
  $(3,2,2)$ & 0.9 & 0.165238 & 0.133333 & 1.239283 \\
  $(3,2,3)$ & 0.9 & 0.170995 & 0.133333 & 1.282464 \\
  $(5,3,2)$ & 0.9 & 0.130506 & 0.086400 & 1.510490 \\
  $(4,2,2)$ & 0.9 & 0.170853 & 0.112500 & 1.518697 \\
  $(2,1,2)$ & 0.9 & 0.170859 & 0.112500 & 1.518750 \\
  $(3,1,2)$ & 0.9 & 0.202500 & 0.133333 & 1.518750 \\
  $(4,1,2)$ & 0.9 & 0.194400 & 0.126562 & 1.535997 \\
  $(5,1,2)$ & 0.9 & 0.180006 & 0.115200 & 1.562552 \\
  $(2,1,3)$ & 0.9 & 0.193473 & 0.112500 & 1.719760 \\
  $(3,1,3)$ & 0.9 & 0.233282 & 0.133333 & 1.749614 \\
  $(2,1,4)$ & 0.9 & 0.198614 & 0.112500 & 1.765455 \\
  $(5,2,2)$ & 0.9 & 0.174555 & 0.086400 & 2.020315 \\
\hline\hline
\end{tabular}
\caption{Lossy candidate-path adaptive benchmarks at \(t=0.9\), used in Fig.~\ref{fig:robustness_comparison}(b). The gains range from \(1.077762\) to \(2.020315\).}
\label{tab:app_adaptive_lossy_values}
\end{table*}

The ideal table (Table \ref{tab:app_adaptive_ideal_values}) shows that the two-path one-absorber value increases from \(1/8\) at one use to \(0.210938\) at two uses and \(0.302714\) at four uses. The three-path one-absorber value increases from \(4/27\) to \(0.250000\) at two uses and \(0.319997\) at three uses. At \((d,k,K)=(5,2,2)\), the gain over the corresponding one-use benchmark is \(2.244701\). The lossy table (Table \ref{tab:app_adaptive_lossy_values}) shows that temporal wiring remains useful at \(t=0.9\) across all solved candidate-path instances.

\subsection{Reference-assisted benchmarks}
\label{app:reference_rail_benchmarks}

We now quantify the effect of a promised empty reference rail across all
computed single-absorber instances. For one absorber hidden uniformly
among $d$ candidate paths, let $U_{d,K}(t)$ denote the optimum over unassisted
$K$-use strategies (acting only on the $d$ candidate paths). Let
$S_{d,K}(t)$ denote the finite balanced Elitzur--Vaidman (EV) scan, and
let $R_{d,K}(t)$ denote the optimum over all $K$-use causal strategies
supplied with one additional rail that is promised not to contain the
absorber. The absorber hypotheses are identical in all three cases. The
reference-assisted optimization enlarges the optical strategy space from
$d$ to $d+1$ paths, but it does not add an absorber hypothesis.

For the finite scan used in this comparison, one balanced EV stage tests
one candidate path against the promised empty rail. With a uniform prior
and per-stage transmissivity $t$, its one- and two-use values are
\begin{equation}
  S_{d,1}(t)=\frac{t}{4d},
  \qquad
  S_{d,2}(t)=\frac{t(1+t)}{4d}.
  \label{eq:app_reference_scan_values}
\end{equation}
The first expression is the probability that the absorber occupies the
single tested candidate and produces the interaction-free dark event.
The second contains the first-stage contribution and the second-stage
contribution after the photon has crossed the preceding empty test.

To compare the strategies both absolutely and relatively, we define
\begin{equation}
\begin{aligned}
  \Delta_{\mathrm{U}}(d,K,t)
  &:=R_{d,K}(t)-U_{d,K}(t),\\
  G_{\mathrm{U}}(d,K,t)
  &:=\frac{R_{d,K}(t)}{U_{d,K}(t)},\\
  \Delta_{\mathrm{S}}(d,K,t)
  &:=R_{d,K}(t)-S_{d,K}(t),\\
  G_{\mathrm{S}}(d,K,t)
  &:=\frac{R_{d,K}(t)}{S_{d,K}(t)}.
\end{aligned}
\label{eq:app_reference_gain_definitions}
\end{equation}
Here $\Delta_{\mathrm{U}}$ and $\Delta_{\mathrm{S}}$ are additive
advantages, while $G_{\mathrm{U}}$ and $G_{\mathrm{S}}$ are
multiplicative gains. Tables~\ref{tab:app_reference_rail_lossless} and
\ref{tab:app_reference_rail_lossy} report all sixteen computed cases.

\begin{table*}[t]
\centering
\begingroup
\scriptsize
\setlength{\tabcolsep}{3.6pt}
\renewcommand{\arraystretch}{1.10}
\begin{tabular}{c c c c c c c c}
\hline\hline
$(d,K)$ &
$U_{d,K}(1)$ &
$S_{d,K}(1)$ &
$R_{d,K}(1)$ &
$\Delta_{\mathrm{U}}$ &
$G_{\mathrm{U}}$ &
$\Delta_{\mathrm{S}}$ &
$G_{\mathrm{S}}$ \\
\hline
$(2,1)$ &
$0.125000$ &
$0.125000$ &
$0.210938$ &
$0.085938$ &
$1.687500$ &
$0.085938$ &
$1.687500$ \\
$(2,2)$ &
$0.210938$ &
$0.250000$ &
$0.347222$ &
$0.136285$ &
$1.646091$ &
$0.097222$ &
$1.388889$ \\
$(3,1)$ &
$0.148148$ &
$0.083333$ &
$0.179558$ &
$0.031410$ &
$1.212019$ &
$0.096225$ &
$2.154701$ \\
$(3,2)$ &
$0.250000$ &
$0.166667$ &
$0.299244$ &
$0.049244$ &
$1.196974$ &
$0.132577$ &
$1.795461$ \\
$(4,1)$ &
$0.140625$ &
$0.062500$ &
$0.154920$ &
$0.014295$ &
$1.101654$ &
$0.092420$ &
$2.478721$ \\
$(4,2)$ &
$0.240002$ &
$0.125000$ &
$0.263568$ &
$0.023566$ &
$1.098191$ &
$0.138568$ &
$2.108547$ \\
$(5,1)$ &
$0.128000$ &
$0.050000$ &
$0.135576$ &
$0.007576$ &
$1.059186$ &
$0.085576$ &
$2.711516$ \\
$(5,2)$ &
$0.222214$ &
$0.100000$ &
$0.235139$ &
$0.012926$ &
$1.058168$ &
$0.135139$ &
$2.351393$ \\
\hline\hline
\end{tabular}
\endgroup
\caption{\label{tab:app_reference_rail_lossless}
Complete lossless reference-rail comparison for one ideal absorber
hidden uniformly among $d$ candidate paths. The columns
$U_{d,K}(1)$, $S_{d,K}(1)$, and $R_{d,K}(1)$ give the unassisted
optimum, the finite balanced EV scan, and the reference-assisted
optimum, respectively. The additive advantages
$\Delta_{\mathrm{U}}$ and $\Delta_{\mathrm{S}}$ and the
multiplicative gains $G_{\mathrm{U}}$ and $G_{\mathrm{S}}$ are defined
in Eq.~\eqref{eq:app_reference_gain_definitions}. For $(d,K)=(2,2)$,
the first three values are the exact quantities $27/128$, $1/4$, and
$25/72$. These values are shown in Fig. \ref{fig:reference_rail_comparison}.}
\end{table*}

\begin{table*}[t]
\centering
\begingroup
\scriptsize
\setlength{\tabcolsep}{3.6pt}
\renewcommand{\arraystretch}{1.10}
\begin{tabular}{c c c c c c c c}
\hline\hline
$(d,K)$ &
$U_{d,K}(0.9)$ &
$S_{d,K}(0.9)$ &
$R_{d,K}(0.9)$ &
$\Delta_{\mathrm{U}}$ &
$G_{\mathrm{U}}$ &
$\Delta_{\mathrm{S}}$ &
$G_{\mathrm{S}}$ \\
\hline
$(2,1)$ &
$0.112500$ &
$0.112500$ &
$0.194908$ &
$0.082408$ &
$1.732516$ &
$0.082408$ &
$1.732516$ \\
$(2,2)$ &
$0.170859$ &
$0.213750$ &
$0.317227$ &
$0.146367$ &
$1.856654$ &
$0.103477$ &
$1.484102$ \\
$(3,1)$ &
$0.133333$ &
$0.075000$ &
$0.163937$ &
$0.030603$ &
$1.229526$ &
$0.088937$ &
$2.185824$ \\
$(3,2)$ &
$0.202500$ &
$0.142500$ &
$0.267750$ &
$0.065250$ &
$1.322220$ &
$0.125250$ &
$1.878945$ \\
$(4,1)$ &
$0.126563$ &
$0.056250$ &
$0.140630$ &
$0.014068$ &
$1.111153$ &
$0.084380$ &
$2.500094$ \\
$(4,2)$ &
$0.194400$ &
$0.106875$ &
$0.232696$ &
$0.038297$ &
$1.197000$ &
$0.125821$ &
$2.177275$ \\
$(5,1)$ &
$0.115200$ &
$0.045000$ &
$0.122693$ &
$0.007493$ &
$1.065045$ &
$0.077693$ &
$2.726516$ \\
$(5,2)$ &
$0.180006$ &
$0.085500$ &
$0.205850$ &
$0.025844$ &
$1.143573$ &
$0.120350$ &
$2.407603$ \\
\hline\hline
\end{tabular}
\endgroup
\caption{\label{tab:app_reference_rail_lossy}
Complete reference-rail comparison at transmissivity $t=0.9$ for the
same single-absorber geometries as in
Table~\ref{tab:app_reference_rail_lossless}. The columns
$U_{d,K}(0.9)$, $S_{d,K}(0.9)$, and $R_{d,K}(0.9)$ have the same
meaning as their lossless counterparts. The reference-assisted optimum
remains above both the unassisted optimum and the finite EV scan in
every computed instance.}
\end{table*}

The complete data establish three consistent features. First,
$R_{d,K}(t)>U_{d,K}(t)$ and $R_{d,K}(t)>S_{d,K}(t)$ in every computed
instance, for $K=1$ and $K=2$ and for both $t=1$ and $t=0.9$. Thus the
promised rail is useful beyond the minimal two-path example, while the
finite scan remains strictly suboptimal when the same rail is optimized
over the full causal class.

Second, the finite scan and the optimized reference-assisted strategy
scale very differently with the number of candidate paths. At fixed
$K=2$, the scan value decreases as
$S_{d,2}(t)=t(1+t)/(4d)$ because only two candidate paths are tested.
The reference-assisted strategy is not restricted to assigning one
candidate to each use; it can distribute, retain, and recombine
amplitudes coherently across the enlarged path space. This is why
$G_{\mathrm{S}}$ grows with $d$ in both tables. In the lossless data,
for example, $G_{\mathrm{S}}$ increases from $1.388889$ at $d=2$ to
$2.351393$ at $d=5$ for $K=2$.

Third, attenuation lowers the absolute values of all three strategies
but increases the relative reference-assisted gains in every matched
$(d,K)$ pair. For $K=2$, $G_{\mathrm{U}}$ changes from $1.646091$ to
$1.856654$ at $d=2$, from $1.196974$ to $1.322220$ at $d=3$, from
$1.098191$ to $1.197000$ at $d=4$, and from $1.058168$ to $1.143573$
at $d=5$ when $t$ is reduced from $1$ to $0.9$. The corresponding
values of $G_{\mathrm{S}}$ also increase in every case. Over the solved
range, loss therefore does not erase the value of the promised rail;
relative to both the unassisted optimum and the finite scan, the
advantage of optimizing that rail coherently becomes more pronounced.

\section{Exact two-use hierarchy for two candidate paths}
\label{app:two_use_exact}

This appendix proves Proposition~\ref{prop:exact_unassisted_two_use}
and Theorem~\ref{thm:exact_reference_assisted_two_use}. We consider one
ideal absorber hidden with equal prior probability in candidate path
$0$ or $1$, and allow two encounters with the absorber region. The
one-shot optimum $1/8$ follows from
Theorem~\ref{thm:one_shot_optimality}. We derive the exact unassisted
two-use optimum $27/128$, evaluate the finite balanced
Elitzur--Vaidman (EV) scan, and prove the exact reference-assisted
optimum $25/72$.

Throughout this appendix, an ancillary or memory system is a tensor
factor carried by the strategy. It does not provide an optical path that
bypasses the absorber region. Such a bypass mode is a different physical
resource and is introduced explicitly only in the reference-assisted
problem as a promised empty rail.

\subsection{Common reduction to binary discrimination}
\label{app:two_use_binary_reduction}

A general two-use strategy may contain arbitrary memory, intermediate
measurements, and feedforward. By recording every intermediate outcome
coherently in the memory, the strategy may be purified without changing
its success probability. We may therefore describe its final surviving
branches by pure, generally subnormalized states.

Let $\ket{e}$ be the normalized final state produced when the
interferometer is empty, and let $\ket{s_0}$ and $\ket{s_1}$ be the
final surviving states under the two absorber hypotheses. A conclusive
measurement effect $M_j$ reporting hypothesis $j\in\{0,1\}$ must never
click when the interferometer is empty. Hence
\begin{equation}
  \bra{e}M_j\ket{e}=0.
  \label{eq:app2u_empty_condition}
\end{equation}
Because $M_j\succeq0$, the left-hand side is
$\|M_j^{1/2}\ket{e}\|^2$. Equation~\eqref{eq:app2u_empty_condition} is
therefore equivalent to
\begin{equation}
  M_j\ket{e}=0.
  \label{eq:app2u_empty_kernel}
\end{equation}
Thus a conclusive effect is insensitive to the component parallel to the
empty-interferometer state. Define the projector onto its orthogonal
complement and the corresponding informative vectors by
\begin{equation}
  Q=\oper{I}-\ket{e}\!\bra e,
  \qquad
  \ket{w_j}=Q\ket{s_j}.
  \label{eq:app2u_informative_vectors}
\end{equation}
The localization problem is therefore the optimal binary discrimination
of the subnormalized states
$\frac12\ket{w_0}\!\bra{w_0}$ and
$\frac12\ket{w_1}\!\bra{w_1}$, where the factor $1/2$ is the prior
probability of each absorber location.

Introduce the total informative norm and the overlap
\begin{align}
  T&=\langle w_0\mid w_0\rangle+\langle w_1\mid w_1\rangle,
  \nonumber\\
  C&=\langle w_0\mid w_1\rangle.
  \label{eq:app2u_TC}
\end{align}
For two subnormalized pure states, the Helstrom formula gives
\begin{equation}
  P_{\rm loc}
  =
  \frac14
  \left[
    T+\sqrt{T^2-4|C|^2}
  \right].
  \label{eq:app2u_helstrom}
\end{equation}
The remainder of each proof is therefore a characterization of the Gram
matrix of $\ket{w_0}$ and $\ket{w_1}$ under the available optical
resources.

\subsection{Exact unassisted two-use optimum}
\label{app:two_use_unassisted}

We first restrict the optical path space to the two candidate paths.
Arbitrary ancillary systems and coherent memory are allowed, but no
additional path is promised to remain empty. For convenience, hypothesis
$j$ labels the candidate path that survives; the absorber occupies the
other path. This is only a relabelling of the two absorber hypotheses.

A general pure input can be expanded in the path basis as
\begin{equation}
  \ket{\Psi}
  =
  \ket{0}\ket{r_0}+\ket{1}\ket{r_1},
  \label{eq:app2u_unassisted_initial}
\end{equation}
where $\ket{r_0}$ and $\ket{r_1}$ are arbitrary, generally
subnormalized vectors in the strategy memory. Normalization of
$\ket{\Psi}$ gives
\begin{equation}
  \langle r_0\mid r_0\rangle+\langle r_1\mid r_1\rangle=1.
  \label{eq:app2u_initial_normalization}
\end{equation}

Under hypothesis $0$, only the path-$0$ component survives the first
encounter; under hypothesis $1$, only the path-$1$ component survives.
Let $V$ be an arbitrary isometry describing everything the strategy does
between the two encounters, including coherent records of any
intermediate measurements. Define
\begin{equation}
  \ket{\chi_0}=V\!\left(\ket{0}\ket{r_0}\right),
  \qquad
  \ket{\chi_1}=V\!\left(\ket{1}\ket{r_1}\right).
  \label{eq:app2u_unassisted_chi}
\end{equation}
The two inputs to $V$ are orthogonal because they occupy different path
states. Isometries preserve inner products, so
\begin{align}
  \langle \chi_0\mid \chi_1\rangle&=0,
  \nonumber\\
  \langle \chi_0\mid \chi_0\rangle
  +\langle \chi_1\mid \chi_1\rangle&=1.
  \label{eq:app2u_unassisted_orthogonality}
\end{align}

Immediately before the second encounter, expand each state in the same
path basis:
\begin{equation}
  \ket{\chi_j}
  =
  \ket{0}\ket{\mu_{0,j}}
  +
  \ket{1}\ket{\mu_{1,j}},
  \qquad j\in\{0,1\}.
  \label{eq:app2u_unassisted_decomposition}
\end{equation}
Here $\ket{\mu_{m,j}}$ is a vector in the memory space. The first index
$m$ specifies the path sent into the second encounter, while the second
index $j$ specifies which path survived the first encounter. These
vectors need not be normalized or mutually orthogonal; their norms and
overlaps contain the most general coherent routing and memory allowed
between the two uses.

Using orthogonality of the path states, the first relation in
Eq.~\eqref{eq:app2u_unassisted_orthogonality} becomes
\begin{equation}
  \langle \mu_{0,0}\mid \mu_{0,1}\rangle
  +
  \langle \mu_{1,0}\mid \mu_{1,1}\rangle
  =0.
  \label{eq:app2u_component_orthogonality}
\end{equation}
This is the only constraint linking the four memory vectors beyond the
overall normalization.

At the second encounter, hypothesis $0$ transmits only path $0$, while
hypothesis $1$ transmits only path $1$. Therefore the two final surviving
states are
\begin{equation}
  \ket{s_0}=\ket{0}\ket{\mu_{0,0}},
  \qquad
  \ket{s_1}=\ket{1}\ket{\mu_{1,1}}.
  \label{eq:app2u_unassisted_final_states}
\end{equation}
When the interferometer is empty, neither encounter removes any path
component. By linearity of $V$, the corresponding final state is
\begin{equation}
  \ket{e}
  =
  \ket{\chi_0}+\ket{\chi_1}.
  \label{eq:app2u_unassisted_empty_state}
\end{equation}
Equation~\eqref{eq:app2u_unassisted_orthogonality} implies
$\langle e\mid e\rangle=1$, so $\ket{e}$ is already normalized.

We now compress the remaining degrees of freedom into four real
parameters. Define
\begin{align}
  x&=\langle \mu_{0,0}\mid \mu_{0,0}\rangle,
  \nonumber\\
  y&=\langle \mu_{1,1}\mid \mu_{1,1}\rangle,
  \nonumber\\
  z&=\langle \mu_{0,1}\mid \mu_{0,0}\rangle=u+iv.
  \label{eq:app2u_unassisted_parameters}
\end{align}
Thus $x$ and $y$ are the squared norms of the two components that survive
both encounters, while $z$ is the overlap between the two memory vectors
routed into path $0$ before the second encounter. It is useful to set
\begin{equation}
  s=x+y,
  \qquad
  \delta=x-y.
  \label{eq:app2u_s_delta}
\end{equation}
Because $x\leq\langle \chi_0\mid \chi_0\rangle$ and
$y\leq\langle \chi_1\mid \chi_1\rangle$, the normalization in
Eq.~\eqref{eq:app2u_unassisted_orthogonality} gives
\begin{equation}
  0\leq s\leq1.
  \label{eq:app2u_s_range}
\end{equation}

Next compute the overlaps of the two surviving states with the empty
state. From Eqs.~\eqref{eq:app2u_unassisted_decomposition} and
\eqref{eq:app2u_unassisted_empty_state},
\begin{equation}
  \alpha:=\langle e\mid s_0\rangle=x+z.
  \label{eq:app2u_alpha}
\end{equation}
For the second hypothesis, Eq.~\eqref{eq:app2u_component_orthogonality}
gives
\begin{equation}
  \langle \mu_{1,0}\mid \mu_{1,1}\rangle=-z^*.
  \label{eq:app2u_second_sector_overlap}
\end{equation}
Consequently,
\begin{equation}
  \beta:=\langle e\mid s_1\rangle=y-z^*.
  \label{eq:app2u_beta}
\end{equation}
The unprojected states $\ket{s_0}$ and $\ket{s_1}$ occupy orthogonal
paths, so $\langle s_0\mid s_1\rangle=0$. Projecting away the empty-state
component therefore gives
\begin{align}
  \langle w_0\mid w_0\rangle
  &=x-|\alpha|^2,
  \nonumber\\
  \langle w_1\mid w_1\rangle
  &=y-|\beta|^2,
  \nonumber\\
  \langle w_0\mid w_1\rangle
  &=-\alpha^*\beta.
  \label{eq:app2u_projected_gram_unassisted}
\end{align}
These three quantities determine the Helstrom value in
Eq.~\eqref{eq:app2u_helstrom}.

We first bound the total informative norm. Using
$\alpha=x+u+iv$ and $\beta=y-u+iv$, one obtains
\begin{align}
  T
  &=x+y-|\alpha|^2-|\beta|^2
  \nonumber\\
  &=s-\frac{s^2}{2}
  -\frac{(\delta+2u)^2}{2}
  -2v^2.
  \label{eq:app2u_T_identity}
\end{align}
The last two terms are nonpositive, and therefore
\begin{equation}
  T\leq\frac{s(2-s)}{2}.
  \label{eq:app2u_T_bound}
\end{equation}

We next bound the square-root term in the Helstrom formula. Substituting
Eqs.~\eqref{eq:app2u_alpha}--\eqref{eq:app2u_beta} into
Eq.~\eqref{eq:app2u_projected_gram_unassisted} and expanding gives the
exact identity
\begin{align}
&s^2(1-s)
-
\left(T^2-4|\alpha\beta|^2\right)
\nonumber\\
&\qquad=
 s\left[(1-s)(\delta+2u)^2+4v^2\right].
  \label{eq:app2u_discriminant_identity}
\end{align}
The right-hand side is nonnegative because $0\leq s\leq1$. Since
$C=-\alpha^*\beta$, this proves
\begin{equation}
  \sqrt{T^2-4|C|^2}
  \leq
  s\sqrt{1-s}.
  \label{eq:app2u_discriminant_bound}
\end{equation}
Combining Eqs.~\eqref{eq:app2u_helstrom},
\eqref{eq:app2u_T_bound}, and
\eqref{eq:app2u_discriminant_bound} yields the scalar upper bound
\begin{equation}
  P_{\rm IF}^{(2)}(2,1)
  \leq
  \frac{s}{8}
  \left[2-s+2\sqrt{1-s}\right].
  \label{eq:app2u_scalar_bound}
\end{equation}

Set $r=\sqrt{1-s}$, so that $s=1-r^2$ and $0\leq r\leq1$. The
right-hand side becomes
\begin{equation}
  f(r)=\frac{(1-r)(1+r)^3}{8}.
  \label{eq:app2u_r_bound}
\end{equation}
Its derivative is
\begin{equation}
  f'(r)=\frac14(1+r)^2(1-2r).
  \label{eq:app2u_r_derivative}
\end{equation}
The unique interior maximum occurs at $r=1/2$, equivalently $s=3/4$.
Substitution into Eq.~\eqref{eq:app2u_r_bound} gives
\begin{equation}
  P_{\rm IF}^{(2)}(2,1)
  \leq
  \frac{27}{128}.
  \label{eq:app2u_unassisted_upper_bound}
\end{equation}

It remains to show that the bound is attainable. No ancilla,
intermediate measurement, or feedforward is required. Prepare
\begin{equation}
  \ket{\psi}
  =
  \frac12\ket{0}+\frac{\sqrt3}{2}\ket{1},
  \label{eq:app2u_optimal_input}
\end{equation}
and apply between the two encounters the path rotation
\begin{equation}
  U=
  \begin{pmatrix}
    \sqrt3/2 & 1/2\\
    -1/2 & \sqrt3/2
  \end{pmatrix}.
  \label{eq:app2u_optimal_rotation}
\end{equation}
For the empty interferometer, the final state is
\begin{equation}
  \ket{e}
  =
  U\ket{\psi}
  =
  \frac{\sqrt3}{2}\ket{0}+\frac12\ket{1}.
  \label{eq:app2u_optimal_empty}
\end{equation}
Choose the normalized direction orthogonal to $\ket{e}$,
\begin{equation}
  \ket{e_\perp}
  =
  \frac12\ket{0}-\frac{\sqrt3}{2}\ket{1}.
  \label{eq:app2u_empty_orthogonal}
\end{equation}
Under hypothesis $1$, path $1$ survives both encounters. After the first
encounter, the state is $(\sqrt3/2)\ket{1}$; after applying $U$ and
surviving the second encounter, it becomes
\begin{equation}
  \ket{s_1}=\frac34\ket{1}.
  \label{eq:app2u_saturating_state}
\end{equation}
Use the conclusive effects
\begin{equation}
  M_1=\ket{e_\perp}\!\bra{e_\perp},
  \qquad
  M_0=0.
  \label{eq:app2u_optimal_measurement}
\end{equation}
The empty-interferometer constraint holds because
$\langle e_\perp\mid e\rangle=0$. The average success probability is
\begin{align}
  P_{\rm loc}
  &=
  \frac12
  \left|\langle e_\perp\mid s_1\rangle\right|^2
  \nonumber\\
  &=
  \frac12
  \left(\frac{3\sqrt3}{8}\right)^2
  =
  \frac{27}{128}.
  \label{eq:app2u_unassisted_saturation}
\end{align}
A shared random swap of paths $0$ and $1$ gives a symmetric
implementation with the same average value. We conclude that
\begin{equation}
  P_{\rm IF}^{(2)}(2,1)
  =
  \frac{27}{128}.
  \label{eq:app2u_unassisted_exact}
\end{equation}
Although the optimization covers the full two-use adaptive class, the
optimum is attained by a fixed coherent circuit. The result therefore
establishes an exact finite-round coherent optimum; measurement-based
adaptivity is not required in this minimal case.

\subsection{Finite balanced two-path EV scan}
\label{app:two_use_scan}

The finite scan uses one additional rail $r$ that is promised empty. At
the first encounter, a balanced EV interferometer tests candidate path
$0$ against $r$. If candidate path $0$ is empty, the photon exits the
bright port with certainty and is routed to the second encounter, which
tests candidate path $1$ against the same rail. If the absorber occupies
the candidate tested at a given encounter, the balanced EV stage
produces a dark-port localization event with probability $1/4$.

For a beam splitter with intensity reflectivity $R$ and transmissivity
$1-R$, the corresponding single-stage success probability is
$R(1-R)\leq1/4$, with equality at $R=1/2$. Thus the balanced stage is
optimal for success probability within a single EV test. Because the
absorber is uniformly distributed between the two candidates and each
candidate is tested once,
\begin{equation}
  P_{\rm scan}^{(2)}(2,1)
  =
  \frac12\frac14
  +
  \frac12\frac14
  =
  \frac14.
  \label{eq:app2u_scan_value}
\end{equation}
The scan is feasible only in the reference-assisted geometry, because
$r$ is an optical mode promised never to contain the absorber.

\subsection{Exact reference-assisted two-use optimum}
\label{app:two_use_reference_assisted}

We now optimize over all two-use causal strategies supplied with the
same promised empty rail. The optical path basis is
$\{\ket r,\ket{0},\ket{1}\}$, while the absorber hypotheses remain only
candidate paths $0$ and $1$.

A general pure input, including arbitrary memory, is
\begin{equation}
  \ket{\Psi}
  =
  \ket r\ket{\eta_r}
  +
  \ket{0}\ket{\eta_0}
  +
  \ket{1}\ket{\eta_1},
  \label{eq:app2u_ref_initial}
\end{equation}
where
\begin{equation}
  p_q=\langle \eta_q\mid \eta_q\rangle,
  \qquad
  p_r+p_0+p_1=1.
  \label{eq:app2u_ref_weights}
\end{equation}
The vectors $\ket{\eta_q}$ are arbitrary memory vectors associated with
the three optical paths.

Order the process hypotheses as $(\varnothing,0,1)$, where
$\varnothing$ denotes the empty interferometer and $j\in\{0,1\}$
denotes an absorber in candidate path $j$. After the first encounter,
the surviving states are
\begin{align}
  \ket{\psi^{(1)}_{\varnothing}}
  &=
  \ket r\ket{\eta_r}
  +\ket{0}\ket{\eta_0}
  +\ket{1}\ket{\eta_1},
  \nonumber\\
  \ket{\psi^{(1)}_0}
  &=
  \ket r\ket{\eta_r}
  +\ket{1}\ket{\eta_1},
  \nonumber\\
  \ket{\psi^{(1)}_1}
  &=
  \ket r\ket{\eta_r}
  +\ket{0}\ket{\eta_0}.
  \label{eq:app2u_ref_first_states}
\end{align}
Their Gram matrix, in the stated hypothesis order, is
\begin{equation}
  G^{(1)}=
  \begin{pmatrix}
    1 & p_r+p_1 & p_r+p_0\\
    p_r+p_1 & p_r+p_1 & p_r\\
    p_r+p_0 & p_r & p_r+p_0
  \end{pmatrix}.
  \label{eq:app2u_first_gram}
\end{equation}
For example, the $(0,1)$ entry equals $p_r$ because the first and second
absorber hypotheses share only the reference-rail component after the
first encounter.

Let $V$ be the arbitrary intermediate isometry. For each hypothesis
$h\in\{\varnothing,0,1\}$, expand its output according to the optical
path entering the second encounter:
\begin{equation}
  V\ket{\psi_h^{(1)}}
  =
  \sum_{q\in\{r,0,1\}}
  \ket q\ket{\nu_{q,h}}.
  \label{eq:app2u_ref_intermediate_decomposition}
\end{equation}
Here $\ket{\nu_{q,h}}$ is a generally subnormalized memory vector. The
index $q$ specifies the path entering the second encounter, and $h$
specifies the absorber hypothesis.

For each path sector $q$, define a $3\times3$ Gram matrix by
\begin{equation}
  (Y_q)_{hh'}
  :=
  \langle \nu_{q,h}\mid \nu_{q,h'}\rangle,
  \qquad
  h,h'\in\{\varnothing,0,1\}.
  \label{eq:app2u_Y_definition}
\end{equation}
Every $Y_q$ is positive semidefinite because it is a Gram matrix. Since
$V$ preserves all inner products and the three path sectors are
orthogonal,
\begin{equation}
  Y_r+Y_0+Y_1=G^{(1)}.
  \label{eq:app2u_gram_decomposition}
\end{equation}
Conversely, any three positive-semidefinite matrices satisfying
Eq.~\eqref{eq:app2u_gram_decomposition} are physically realizable. One
chooses vector families with Gram matrices $Y_r$, $Y_0$, and $Y_1$,
places them in mutually orthogonal path sectors, and takes their direct
sum. The resulting hypothesis vectors have Gram matrix $G^{(1)}$, so an
isometry exists from the actual post-first-use states to these vectors on
their span. Equation~\eqref{eq:app2u_gram_decomposition} is therefore an
exact parametrization of the intermediate strategy, not a relaxation.

During the second encounter, a component in the reference sector
survives every hypothesis. A component in candidate path $0$ is removed
only under hypothesis $0$, while a component in candidate path $1$ is
removed only under hypothesis $1$. In the hypothesis order
$(\varnothing,0,1)$, define the corresponding masks
\begin{equation}
  D_0=\operatorname{diag}(1,0,1),
  \qquad
  D_1=\operatorname{diag}(1,1,0).
  \label{eq:app2u_survival_masks}
\end{equation}
The final Gram matrix is therefore
\begin{equation}
  G^{(2)}
  =
  Y_r+D_0Y_0D_0+D_1Y_1D_1.
  \label{eq:app2u_second_gram}
\end{equation}
The empty-empty entry remains equal to one because no path is removed
under the empty hypothesis and
$Y_r+Y_0+Y_1=G^{(1)}$.

We may restrict the optimization to real Gram matrices. Indeed, all
process operators and all constraints are real in the path basis. Given
any feasible strategy and its complex-conjugate strategy, implement the
two in orthogonal flag sectors with equal weight and use the
corresponding block-diagonal measurement. The objective value is
unchanged, while the flagged hypothesis states have Gram matrix
$(G+\overline G)/2=\operatorname{Re}G$. Hence an optimum exists with
real $G^{(2)}$.

Let $\ket{e}$, $\ket{s_0}$, and $\ket{s_1}$ denote the final states
whose Gram matrix is $G^{(2)}$. Define their empty-state overlaps by
\begin{equation}
  \gamma_0=G^{(2)}_{\varnothing0}=\langle e\mid s_0\rangle,
  \qquad
  \gamma_1=G^{(2)}_{\varnothing1}=\langle e\mid s_1\rangle.
  \label{eq:app2u_gamma}
\end{equation}
Since $G^{(2)}_{\varnothing\varnothing}=1$, projection orthogonal to
$\ket{e}$ gives the informative Gram matrix
\begin{equation}
  W=
  \begin{pmatrix}
    A & C\\
    C & B
  \end{pmatrix},
  \label{eq:app2u_projected_W}
\end{equation}
with
\begin{equation}
\begin{aligned}
  A&=G^{(2)}_{00}-\gamma_0^2,\\
  B&=G^{(2)}_{11}-\gamma_1^2,\\
  C&=G^{(2)}_{01}-\gamma_0\gamma_1.
\end{aligned}
  \label{eq:app2u_projected_gram_ref}
\end{equation}
These are simply the inner products
$A=\langle w_0\mid w_0\rangle$,
$B=\langle w_1\mid w_1\rangle$, and
$C=\langle w_0\mid w_1\rangle$ after subtracting the empty-state component.
Set
\begin{equation}
  T=A+B,
  \qquad
  X=T+2C,
  \qquad
  Y=T-2C.
  \label{eq:app2u_XY}
\end{equation}
Because $W\succeq0$, one has $|C|\leq(A+B)/2$, and therefore
$X,Y\geq0$. Moreover,
$XY=T^2-4C^2$. Equation~\eqref{eq:app2u_helstrom} can thus be written as
\begin{equation}
  P_{\rm loc}
  =
  \frac18\left(\sqrt X+\sqrt Y\right)^2.
  \label{eq:app2u_helstrom_XY}
\end{equation}

\subsubsection{A physical support family}

Equation~\eqref{eq:app2u_helstrom_XY} reduces the optimization to the
pair $(T,C)$, but not every positive $2\times2$ Gram matrix is compatible
with two encounters and one promised empty reference path. We now derive
the required restriction directly from the path-sector Gram matrices.

Consider a linear support
\begin{equation}
  aT+bC\leq c.
  \label{eq:app2u_general_support}
\end{equation}
A sufficient proof of Eq.~\eqref{eq:app2u_general_support} is an exact
decomposition of its gap into quadratic forms that are nonnegative for
every physical strategy. Let
\begin{equation}
  \gamma=
  \begin{pmatrix}\gamma_0\\ \gamma_1\end{pmatrix},
  \qquad
  \mathbf 1=
  \begin{pmatrix}1\\1\end{pmatrix}.
  \label{eq:app2u_gamma_vector}
\end{equation}
We seek an identity of the form
\begin{align}
  c-aT-bC
  ={}&
  (\gamma-\ell\mathbf 1)^{T}M(\gamma-\ell\mathbf 1)
  \nonumber\\
  &+\operatorname{Tr}(K_rY_r)
  +\operatorname{Tr}(K_0Y_0)
  +\operatorname{Tr}(K_1Y_1),
  \label{eq:app2u_positive_gap_ansatz}
\end{align}
where $M,K_r,K_0,K_1\succeq0$. The exchange symmetry of the two
candidate paths allows the coefficient matrices to be chosen as
\begin{equation}
  M=
  \begin{pmatrix}m&n\\n&m\end{pmatrix},
  \qquad
  K_1=SK_0S,
  \qquad
  S=
  \begin{pmatrix}
    1&0&0\\
    0&0&1\\
    0&1&0
  \end{pmatrix},
  \label{eq:app2u_exchange_matrices}
\end{equation}
with
\begin{equation}
  K_r=
  \begin{pmatrix}
    \rho_0&\rho_1&\rho_1\\
    \rho_1&\rho_2&\rho_3\\
    \rho_1&\rho_3&\rho_2
  \end{pmatrix}.
  \label{eq:app2u_Kr_general}
\end{equation}

The coefficients in Eq.~\eqref{eq:app2u_positive_gap_ansatz} are not
chosen independently. Substitute
$Y_1=G^{(1)}-Y_r-Y_0$ and $p_1=1-p_r-p_0$, and then use
Eqs.~\eqref{eq:app2u_first_gram}, \eqref{eq:app2u_second_gram}, and
\eqref{eq:app2u_projected_gram_ref}. Matching the coefficients of the
independent entries of $Y_r$, $Y_0$, $p_r$, and $p_0$ gives
\begin{equation}
  a=m,
  \qquad
  b=2n,
  \label{eq:app2u_matching_ab}
\end{equation}
\begin{equation}
  u=-\frac{\rho_2}{2}-\rho_3-\frac{m}{2}-n,
  \qquad
  \rho_1=u+\ell(m+n),
  \label{eq:app2u_matching_u}
\end{equation}
\begin{equation}
  K_0=
  \begin{pmatrix}
    \rho_0&u&\rho_1\\
    u&\rho_2+m&\rho_3+n\\
    \rho_1&\rho_3+n&\rho_2
  \end{pmatrix},
  \label{eq:app2u_matching_K0}
\end{equation}
and
\begin{equation}
  c=2\ell^2(m+n)+\rho_0-2\rho_3-2n.
  \label{eq:app2u_matching_c}
\end{equation}
These relations are simply the coefficient conditions under which the
right-hand side of Eq.~\eqref{eq:app2u_positive_gap_ansatz} equals the
desired gap for every feasible strategy.

To obtain a sharp support, it is enough to exhibit one positive
rank-one realization of the matched matrices. Sharpness will be fixed
independently by an attaining physical strategy below. Introduce a
parameter $\zeta>1$ and take
\begin{equation}
\begin{aligned}
  K_r&=v_r^{(\zeta)}v_r^{(\zeta)T},
  &v_r^{(\zeta)}&=
  \begin{pmatrix}(\zeta+2)/2\\-1\\-1\end{pmatrix},\\
  K_0&=v_0^{(\zeta)}v_0^{(\zeta)T},
  &v_0^{(\zeta)}&=
  \begin{pmatrix}(\zeta+2)/2\\-\zeta\\-1\end{pmatrix},\\
  K_1&=v_1^{(\zeta)}v_1^{(\zeta)T},
  &v_1^{(\zeta)}&=
  \begin{pmatrix}(\zeta+2)/2\\-1\\-\zeta\end{pmatrix}.
\end{aligned}
  \label{eq:app2u_support_vectors}
\end{equation}
Comparing these factorizations with
Eqs.~\eqref{eq:app2u_matching_u}--\eqref{eq:app2u_matching_K0}
fixes the remaining coefficients successively:
\begin{equation}
  n=\zeta-1,
  \qquad
  m=\zeta^2-1,
  \qquad
  \ell=\frac12,
  \qquad
  c=\frac{\zeta(3\zeta-2)}{4}.
  \label{eq:app2u_support_coefficients}
\end{equation}
Substitution into Eq.~\eqref{eq:app2u_positive_gap_ansatz} yields the
exact identity
\begin{widetext}
\begin{align}
&\frac{\zeta(3\zeta-2)}{4}
-
\left[(\zeta^2-1)T+2(\zeta-1)C\right]
\nonumber\\
&\quad=
\left(\gamma-\frac12\mathbf 1\right)^T
\begin{pmatrix}
  \zeta^2-1&\zeta-1\\
  \zeta-1&\zeta^2-1
\end{pmatrix}
\left(\gamma-\frac12\mathbf 1\right)
+
(v_r^{(\zeta)})^TY_rv_r^{(\zeta)}
+
(v_0^{(\zeta)})^TY_0v_0^{(\zeta)}
+
(v_1^{(\zeta)})^TY_1v_1^{(\zeta)}.
  \label{eq:app2u_exact_support_identity}
\end{align}
\end{widetext}
The two eigenvalues of the $2\times2$ matrix are
$\zeta(\zeta-1)$ and $(\zeta-1)(\zeta+2)$, while every remaining term
is nonnegative because $Y_r,Y_0,Y_1\succeq0$. Hence, for every
$\zeta>1$,
\begin{equation}
  (\zeta+1)T+2C
  \leq
  \frac{\zeta(3\zeta-2)}{4(\zeta-1)}.
  \label{eq:app2u_support_TC}
\end{equation}
Using $T=(X+Y)/2$ and $C=(X-Y)/4$, this is equivalently
\begin{equation}
  (\zeta+2)X+\zeta Y
  \leq
  \frac{\zeta(3\zeta-2)}{2(\zeta-1)}.
  \label{eq:app2u_support_XY}
\end{equation}

\subsubsection{Optimization of the support parameter}

Set $u=\sqrt X$ and $v=\sqrt Y$. Weighted Cauchy--Schwarz, with the
same coefficients as in Eq.~\eqref{eq:app2u_support_XY}, gives
\begin{align}
  (u+v)^2
  &\leq
  \left[(\zeta+2)u^2+\zeta v^2\right]
  \left(\frac1{\zeta+2}+\frac1\zeta\right)
  \nonumber\\
  &\leq
  \frac{(3\zeta-2)(\zeta+1)}{(\zeta-1)(\zeta+2)}.
  \label{eq:app2u_weighted_CS}
\end{align}
Therefore
\begin{equation}
  P_{\rm loc}
  \leq
  F(\zeta)
  :=
  \frac{(3\zeta-2)(\zeta+1)}
       {8(\zeta-1)(\zeta+2)}.
  \label{eq:app2u_Fzeta}
\end{equation}
Its derivative is
\begin{equation}
  F'(\zeta)
  =
  \frac{\zeta(\zeta-4)}
       {4(\zeta-1)^2(\zeta+2)^2}.
  \label{eq:app2u_Fzeta_derivative}
\end{equation}
Thus $F$ decreases on $1<\zeta<4$, increases for $\zeta>4$, and has a
unique minimum at $\zeta=4$. Equations~\eqref{eq:app2u_support_TC} and
\eqref{eq:app2u_support_XY} become
\begin{equation}
  5T+2C\leq\frac{10}{3},
  \qquad
  3X+2Y\leq\frac{10}{3},
  \label{eq:app2u_optimal_support}
\end{equation}
and therefore
\begin{equation}
  P_{\rm loc}\leq F(4)=\frac{25}{72}.
  \label{eq:app2u_ref_upper_bound}
\end{equation}
The equality conditions determine the contact point. Equality in the
weighted Cauchy--Schwarz step requires $3\sqrt X=2\sqrt Y$; together
with $3X+2Y=10/3$, this gives
\begin{equation}
  X=\frac49,
  \qquad
  Y=1.
  \label{eq:app2u_contact_point}
\end{equation}
Equivalently,
\begin{equation}
  \frac{5}{48}(3X+2Y)
  -\frac18(\sqrt X+\sqrt Y)^2
  =
  \frac{(3\sqrt X-2\sqrt Y)^2}{48}
  \geq0.
  \label{eq:app2u_contact_identity}
\end{equation}
This shows explicitly how the physical support and the discrimination
objective meet at the optimum.

\subsubsection{Explicit attaining strategy}

We now specify a physical strategy attaining the bound. Prepare
\begin{equation}
  \ket{\Psi_\star}
  =
  \frac{1}{\sqrt2}\ket r
  +\frac12\ket0
  +\frac12\ket1,
  \label{eq:app2u_optimal_input_ref}
\end{equation}
so that
\begin{equation}
  p_r=\frac12,
  \qquad
  p_0=p_1=\frac14.
  \label{eq:app2u_optimal_weights}
\end{equation}
The first-use Gram matrix is then
\begin{equation}
  G^{(1)}=
  \begin{pmatrix}
    1&3/4&3/4\\
    3/4&3/4&1/2\\
    3/4&1/2&3/4
  \end{pmatrix}.
  \label{eq:app2u_optimal_first_gram}
\end{equation}
Choose the path-sector Gram matrices
\begin{equation}
  Y_r=
  \begin{pmatrix}
    2/27&1/9&1/9\\
    1/9&2/9&1/9\\
    1/9&1/9&2/9
  \end{pmatrix},
  \label{eq:app2u_Yr}
\end{equation}
\begin{equation}
  Y_0=
  \begin{pmatrix}
    25/54&1/4&7/18\\
    1/4&5/36&7/36\\
    7/18&7/36&7/18
  \end{pmatrix},
  \label{eq:app2u_Y0}
\end{equation}
and
\begin{equation}
  Y_1=
  \begin{pmatrix}
    25/54&7/18&1/4\\
    7/18&7/18&7/36\\
    1/4&7/36&5/36
  \end{pmatrix}.
  \label{eq:app2u_Y1}
\end{equation}
They satisfy $Y_r+Y_0+Y_1=G^{(1)}$. Their positivity is seen directly
from the exact factorizations $Y_q=L_q^TL_q$, where
\begin{equation}
  L_r=
  \begin{pmatrix}
    \sqrt6/9&\sqrt6/6&\sqrt6/6\\
    0&\sqrt2/6&-\sqrt2/6
  \end{pmatrix},
  \label{eq:app2u_Lr}
\end{equation}
\begin{equation}
  L_0=
  \begin{pmatrix}
    5\sqrt6/18&3\sqrt6/20&7\sqrt6/30\\
    0&\sqrt{14}/60&-\sqrt{14}/15
  \end{pmatrix},
  \label{eq:app2u_L0}
\end{equation}
and
\begin{equation}
  L_1=
  \begin{pmatrix}
    5\sqrt6/18&7\sqrt6/30&3\sqrt6/20\\
    0&\sqrt{14}/15&-\sqrt{14}/60
  \end{pmatrix}.
  \label{eq:app2u_L1}
\end{equation}
The three columns of $L_q$ are the memory vectors
$\ket{\nu_{q,\varnothing}}$, $\ket{\nu_{q,0}}$, and
$\ket{\nu_{q,1}}$ in the second-use path sector $q$. Consequently,
these matrices already give an exact Gram realization of the
intermediate operation.

For completeness, the attaining isometry can be written explicitly.
Let the memory be two dimensional with basis
$\{\ket{0_M},\ket{1_M}\}$. In the input basis
$(\ket r,\ket0,\ket1)$ and output basis
\begin{equation*}
  (\ket r\ket{0_M},\ket r\ket{1_M},
   \ket0\ket{0_M},\ket0\ket{1_M},
   \ket1\ket{0_M},\ket1\ket{1_M}),
\end{equation*}
take
\begin{widetext}
\begin{equation}
  V_\star=
  \begin{pmatrix}
    4\sqrt3/9&-\sqrt6/9&-\sqrt6/9\\
    0&-\sqrt2/3&\sqrt2/3\\
    19\sqrt3/90&23\sqrt6/90&4\sqrt6/45\\
    -\sqrt7/10&-\sqrt{14}/30&2\sqrt{14}/15\\
    19\sqrt3/90&4\sqrt6/45&23\sqrt6/90\\
    \sqrt7/10&-2\sqrt{14}/15&\sqrt{14}/30
  \end{pmatrix},
  \qquad
  V_\star^\dagger V_\star=I_3.
  \label{eq:app2u_explicit_isometry}
\end{equation}
\end{widetext}
Applied after the first surviving encounter, $V_\star$ produces the
path-sector vectors in Eqs.~\eqref{eq:app2u_Lr}--\eqref{eq:app2u_L1}.
The second encounter then acts on the output path while the memory is
retained. This is a genuinely two-round adaptive strategy in the
quantum-comb sense: the state entering the second encounter is generated
coherently from the branch that survived the first encounter. The
adaptation is carried by quantum memory and does not require an
intermediate measurement or classical feedforward.

Applying the second-use masks gives
\begin{equation}
  G^{(2)}=
  \begin{pmatrix}
    1&1/2&1/2\\
    1/2&11/18&1/9\\
    1/2&1/9&11/18
  \end{pmatrix}.
  \label{eq:app2u_optimal_second_gram}
\end{equation}
Thus $\gamma_0=\gamma_1=1/2$, and projection orthogonal to the empty
state gives
\begin{equation}
  W=
  \begin{pmatrix}
    13/36&-5/36\\
    -5/36&13/36
  \end{pmatrix}.
  \label{eq:app2u_optimal_projected_gram}
\end{equation}
Hence
\begin{equation}
  T=\frac{13}{18},
  \qquad
  C=-\frac{5}{36},
  \qquad
  X=\frac49,
  \qquad
  Y=1.
  \label{eq:app2u_optimal_T_C_X_Y}
\end{equation}
The support identity is saturated at $\zeta=4$ because
\begin{equation}
\begin{aligned}
  Y_r\begin{pmatrix}3\\-1\\-1\end{pmatrix}&=0,
  &Y_0\begin{pmatrix}3\\-4\\-1\end{pmatrix}&=0,
  &Y_1\begin{pmatrix}3\\-1\\-4\end{pmatrix}&=0,
\end{aligned}
  \label{eq:app2u_support_saturation}
\end{equation}
and $\gamma_0=\gamma_1=1/2$.

It remains only to specify the final readout. Define the normalized
symmetric and antisymmetric informative directions
\begin{equation}
  \ket{\xi_+}
  =\frac{\ket{w_0}+\ket{w_1}}{\sqrt X},
  \qquad
  \ket{\xi_-}
  =\frac{\ket{w_0}-\ket{w_1}}{\sqrt Y}.
  \label{eq:app2u_xi_pm}
\end{equation}
They are orthonormal and orthogonal to the empty-interferometer state.
At the point in Eq.~\eqref{eq:app2u_optimal_T_C_X_Y},
\begin{equation}
  \ket{w_0}=\frac13\ket{\xi_+}+\frac12\ket{\xi_-},
  \qquad
  \ket{w_1}=\frac13\ket{\xi_+}-\frac12\ket{\xi_-}.
  \label{eq:app2u_w_decomposition}
\end{equation}
The Helstrom measurement is
\begin{equation}
  \ket{m_0}=\frac{\ket{\xi_+}+\ket{\xi_-}}{\sqrt2},
  \qquad
  \ket{m_1}=\frac{\ket{\xi_+}-\ket{\xi_-}}{\sqrt2},
  \label{eq:app2u_m_pm}
\end{equation}
with
\begin{equation}
\begin{aligned}
  M_0&=\ket{m_0}\!\bra{m_0},
  \qquad
  M_1=\ket{m_1}\!\bra{m_1},\\
  M_?&=I-M_0-M_1.
\end{aligned}
  \label{eq:app2u_final_povm}
\end{equation}
Because $\ket{m_0},\ket{m_1}\perp\ket e$, the empty-interferometer
condition is exact. Moreover,
\begin{equation}
  |\langle m_0|w_0\rangle|^2
  =
  |\langle m_1|w_1\rangle|^2
  =
  \frac12\left(\frac13+\frac12\right)^2
  =
  \frac{25}{72}.
  \label{eq:app2u_measurement_success}
\end{equation}
Averaging the two equal contributions with the equal priors therefore
gives
\begin{equation}
  P_{\rm IF,ref}^{(2)}(2,1)=\frac{25}{72}.
  \label{eq:app2u_ref_exact}
\end{equation}
This attains Eq.~\eqref{eq:app2u_ref_upper_bound} and completes the proof
of Theorem~\ref{thm:exact_reference_assisted_two_use}.

\subsection{Exact hierarchy and resource interpretation}

The one-shot theorem gives
$P_{\rm IF}^{(1)}(2,1)=1/8$. Combining this value with the two exact
optima and the finite balanced scan gives
\begin{equation}
  \frac18
  <
  \frac{27}{128}
  <
  \frac14
  <
  \frac{25}{72}.
  \label{eq:app2u_exact_hierarchy}
\end{equation}
The first inequality is the gain from a second encounter with the
absorber region. The second reflects the addition of a promised empty
rail, which makes the finite scan feasible. The final inequality proves
that sequential scanning is not the optimal use of that rail. This
completes the proofs of
Proposition~\ref{prop:exact_unassisted_two_use} and
Theorem~\ref{thm:exact_reference_assisted_two_use}.

\section{Loss comparison with sequential Zeno localization}
\label{app:loss_comparison}

This appendix derives the two loss formulas used in Fig.~\ref{fig:robustness_comparison}. The comparison is intentionally simple: one coherent multipath test is compared with a sequential Zeno scan under the same per-pass transmissivity parameter \(t\).

For one ideal absorber, the lossy multipath pass has dark probability
\begin{equation}
  P_s^{\rm HD}(t)=\frac{t(d-1)}{d^2},
  \label{eq:app_hd_dark_loss}
\end{equation}
and bright probability
\begin{equation}
  P_b^{\rm HD}(t)=\frac{t(d-1)^2}{d^2} .
  \label{eq:app_hd_bright_loss}
\end{equation}
Bright events are inconclusive and can be recycled. The probability of eventually reaching a dark branch before loss or absorption is the geometric sum
\begin{equation}
  P_s^{\rm HD}(t)
  \sum_{r=0}^\infty [P_b^{\rm HD}(t)]^r
  =
  \frac{t(d-1)}{d^2-t(d-1)^2} .
  \label{eq:app_hd_recycling}
\end{equation}
The dark-branch state is the same regular-simplex code as in the lossless case, so the optimal probability of identifying the occupied path conditioned on a dark branch is \((d-1)/d\). Thus
\begin{equation}
  \lambda_d^{\rm HD}(t)
  =
  \frac{d-1}{d}\,
  \frac{t(d-1)}{d^2-t(d-1)^2} .
  \label{eq:app_hd_loss_localization}
\end{equation}
At \(t=1\), this becomes \((d-1)^2/[d(2d-1)]\), the lossless localization efficiency.

For a sequential Zeno scan, suppose each candidate location is tested with \(m\) weak recursions. If the absorber is at location \(q\), the photon first survives \(q-1\) empty tests and then produces the Zeno interaction-free success at the occupied location. The transmissivity factor is \(t^{mq}\), and the Zeno success factor for the occupied test is \(\cos^{2m}(\pi/2m)\). Averaging over a uniform absorber position gives
\begin{align}
  \lambda_d^{\rm Zeno}(t,m)
  &=
  \frac1d\sum_{q=1}^d
  t^{mq}\cos^{2m}\!\left(\frac{\pi}{2m}\right)
  \nonumber\\
  &=
  \frac{t^m}{d}
  \cos^{2m}\!\left(\frac{\pi}{2m}\right)
  \frac{1-t^{md}}{1-t^m},
  \label{eq:app_zeno_loss_localization}
\end{align}
with the \(t=1\) value understood by continuity. For \(d=6\) and \(m=10\), the two curves cross at \(t\simeq0.9746\). At \(t=0.95\), Eq.~\eqref{eq:app_hd_loss_localization} gives \(0.323\) and Eq.~\eqref{eq:app_zeno_loss_localization} gives \(0.185\). At \(t=0.90\), the corresponding values are \(0.278\) and \(0.0695\).

\section{Scalar coherent attenuation as a special case}
\label{app:scalar_attenuation}

A scalar coherent attenuation model is obtained by taking one Kraus operator
\begin{equation}
  K_a(\tau)=I-(1-\tau)\ket a\bra a,
  \qquad |\tau|\le1 .
  \label{eq:app_scalar_attenuator}
\end{equation}
The ideal absorber is \(\tau=0\), and the empty device is \(\tau=1\). For the balanced single-pass protocol, the dark component is proportional to \(1-\tau\). Hence
\begin{equation}
  P_s(\tau)
  =
  \frac{|1-\tau|^2(d-1)}{d^2} .
  \label{eq:app_scalar_dark_rate}
\end{equation}
For real positive \(\tau=\sqrt r\), this becomes
\begin{equation}
  P_s(r)
  =
  \frac{(1-\sqrt r)^2(d-1)}{d^2} .
  \label{eq:app_scalar_opacity}
\end{equation}
This example shows how finite coherent transparency suppresses the dark-port rate. It is not the general absorber model; the general model is the calibrated no-click operation \(\mathcal S_j^{\rm exp}\).

\bibliographystyle{apsrev4-2}
\bibliography{IFM_bib}

\end{document}